\documentclass[letterpaper, 10pt]{article}
\usepackage{geometry}
\usepackage{amsmath,amssymb,cite,url}
\usepackage{graphicx}
\usepackage{color}
\usepackage{acronym}

\newcommand{\argmin}[0]{\text{argmin}} 

\newcommand{\eg}[0]{\textit{e.g.} }
\newcommand{\ie}[0]{\textit{i.e.} }
\newcommand{\cf}[0]{\textit{cf.} }

\newcommand{\nm}[0]{[n,m]}

\newcommand{\te}[0]{T_s}
\newcommand{\fs}[0]{F_s}

\acrodef{bss}[BSS]{Blind Source Separation}
\acrodef{bass}[BASS]{Blind Audio Source Separation}
\acrodef{iss}[ISS]{Informed Source Separation}
\acrodef{cnn}[CNN]{Convolutional Neural Networks}
\acrodef{duet}[DUET]{Degenerate Unmixing Estimation Technique}
\acrodef{kam}[KAM]{Kernel Additive Modeling}
\acrodef{ica}[ICA]{Independent Component Analysis}
\acrodef{irmfsp}[IRMFSP]{Inertia Ratio Maximization using Features Space Projection}
\acrodef{hpss}[HPSS]{Harmonic Percussive Source Separation}
\acrodef{mir}[MIR]{Music Information Retrieval}
\acrodef{rpca}[RPCA]{Robust Principal Component Analysis}
\acrodef{stft}[STFT]{Short-Time Fourier Transform}
\acrodef{casa}[CASA]{Computational Auditory Scene Analysis}
\acrodef{nmf}[NMF]{Non-negative Matrix Factorization}
\acrodef{lda}[LDA]{Linear Discriminant Analysis}
\acrodef{ops}[OPS]{Overall Perceptual Score}
\acrodef{svm}[SVM]{Support Vector Machines}
\acrodef{snr}[SNR]{Signal-to-Noise Ratio}
\acrodef{sct}[SCT]{Scattering Transform}
\acrodef{ttb}[TTB]{Timbre ToolBox}

\acrodef{vtmr}[VTMR]{Voice-to-Music Ratio}
\acrodef{sar}[SAR]{Signal-to-Artifact Ratio}
\acrodef{sdr}[SDR]{Signal-to-Distortion Ratio}
\acrodef{sir}[SIR]{Signal-to-Interference Ratio}

\acrodef{rqf}[RQF]{Reconstruction Quality Factor}
\acrodef{se}[SE]{Spectral Entropy}
\acrodef{bw}[BW]{Frequency Bandwidth}
\acrodef{bnl}[BNL]{Background Noise Level}
\acrodef{tfr}[TFR]{Time-Frequency Representation}
\acrodef{mse}[MSE]{Mean Squared Error}
\acrodef{mfcc}[MFCC]{Mel Frequency Cepstral Coefficients}

\usepackage[vlined,ruled]{algorithm2e}
\usepackage[]{subfigure}
\usepackage{url}
\begin{document}
\title{Single-Channel Blind Source Separation for Singing Voice Detection: A Comparative Study}

\author{Dominique~Fourer and Geoffroy Peeters}

\maketitle

\begin{abstract}
We propose a novel unsupervised singing voice detection method which use 
single-channel Blind Audio Source Separation (BASS) algorithm as a preliminary step.
To reach this goal, we investigate three promising BASS approaches
which operate through a morphological filtering of the analyzed mixture spectrogram.
The contributions of this paper are manyfold.
First, the investigated BASS methods are reworded with the same formalism
and we investigate their respective hyperparameters by numerical simulations.
Second, we propose an extension of the KAM method for which we propose a novel
training algorithm used to compute a source-specific kernel from a given isolated source signal.
Second, the BASS methods are compared together in terms of source separation accuracy and in
terms of singing voice detection accuracy when they are used in our new singing voice detection framework.
Finally, we do an exhaustive singing voice detection evaluation for which we compare 
both supervised and unsupervised singing voice detection methods.
Our comparison explores different combination of the proposed BASS methods with new features such 
as the new proposed KAM features and the scattering transform through a machine learning framework
and also considers convolutional neural networks methods.
\end{abstract}

\section{Introduction}\label{sec:introduction}

Audio source separation aims at recovering the isolated signals of each source (\ie each instrumental part)
which composes an observed mixture \cite{comon2010handbook, bss2}.
Although humans can easily recognize the different sound entities which are active at each time instant, this task 
remains challenging when it has to be automatically completed by an unsupervised algorithm.
Mathematically speaking, \ac{bass} is an ``ill-posed problem'' in the sense of Hadamard \cite{idier2013bayesian}, 
however it remains intensively studied since many decades \cite{comon2010handbook,vincent2007first,Liutkus2016,virtanen2016,CreagerSBD16}.
In fact, \ac{bass} is full of interest because it can find many applications such as music remixing (karaoke, re-spatialization, source manipulation),
and signal enhancement (denoising). % and most of the polyphonic music processing applications \cite{pratzlich2015kernel,yela2016interference}.
Thus, \ac{bass} can directly be used as a part of a signal detection method (\ie singing voice), in relation with the source separation model.
This study, addresses the single-channel blind case, when several sources $s_i$ ($i\in [1,I]$, with $I \geq 2$) are present 
in a unique instantaneous mixture $x$ expressed as:
%--------------------------
\begin{equation}
 x(t) = \displaystyle\sum_{i=1}^I s_i(t).
 \label{eq:mixmodel}
\end{equation}
%--------------------------
%
%% TO IMPROVE
Despite the simplicity of the mixture model of Eq.~\eqref{eq:mixmodel}, this configuration is more challenging to solve
than multi-channel mixtures. In fact, multi-channel methods such as \cite{bss_duet,bss2} require at least 2 distinct observed mixtures
with a sufficient orthogonality in the time-frequency plane between the sources, to provide satisfying separation results.
As we address the underdetermined case (where the number of sources is greater than the number
of observations), \ac{ica} methods can neither be directly used \cite{comon2010handbook}.
Moreover, methods inspired by \ac{casa} \cite{casa}, such as \cite{creager2016,Liutkus2016,fourergretsi17},
are often not robust enough for processing real-world music mixtures and should be addressed through an \ac{iss}
framework using side-information in a coder-decoder scheme as proposed in \cite{fourer2013}.

For all these reasons, we focus on another class of robust \ac{bass} methods based on time-frequency representation morphological filtering.
These methods assume that the foreground voice and the instrumental music background have significantly different time-frequency regularities
which can be exploited to assign each time-frequency point to a source.
To illustrate this idea, vertical lines can be observed in a drum set spectrogram due the spectral regularities at each instant,
contrarily to an harmonic source which has horizontal lines due to the regularities over time of each active frequency (\ie the partials). 
A recent comparative study \cite{lehner2015monaural} leads us to three very promising approaches which can be summarized as follows.
%\begin{enumerate}
\vspace{0.1cm}

1) Total variation approach proposed by Jeong and Lee \cite{jeong2014vocal}, aims at minimizing a convex auxiliary function, 
related to the temporal continuity (for harmonic sources), the spectral continuity (for percussive sounds) and the sparsity 
for the leading singing voice. The solutions provides estimates of the spectrogram of each source.

2) \ac{rpca}\cite{candes2011robust} is used for voice/music separation in \cite{huang2012singing}. This technique
decomposes the mixture spectrogram into two matrices: a low rank matrix associated to the spectrogram of the repetitive musical background (the accompaniment),
and a sparse matrix associated to the lead instrument which plays the melody.

3) \ac{kam} as formalized in \cite{liutkus2014}, unifies several \ac{bass} approaches into the same framework:
REPET \cite{rafii2013repeating} and \ac{hpss} through median filtering \cite{fitzgerald2010harmonic}.
Both methods use the source-specific regularities in their time-frequency representations to compute a source separation mask.
Hence, each source is characterized by a kernel which models the vicinity of each time-frequency point in a spectrogram. 
This allows to estimate each source using a median filter based on its specific kernel.
This idea was extended through other source-specific kernels in \cite{liutkus2014,liutkus2015,kim2015,cho2015singing}
and in the present paper.
%\end{enumerate}
%\vspace{0.1cm}

Thus, the purpose of this work is first to unify these \ac{bass} methods into the same framework to segregate a monaural mixture into 3 components
corresponding to the percussive part, the harmonic background and the singing voice.
Second, we introduce a new unsupervised singing voice detection method which can use any \ac{bass} method as a preprocessing step.
Finally, the \ac{bass} methods are compared together in terms of separation quality and in terms of singing voice detection accuracy.
Our evaluation also considers a comparison with supervised state-of-the-art singing voice detection methods such as \cite{schluter2016learning} 
which uses deep \ac{cnn}.

This paper is organized as follows. In Section \ref{sec:methods}, we shortly describe the proposed \ac{bass} methods
with an extension of the \ac{kam} method for source-specific kernel training. In Section \ref{sec:voice}, we introduce our framework for singing voice detection
based on \ac{bass}. In Section \ref{sec:results}, comparative results for source separation and singing voice detection are presented.
Finally, conclusion and future works are discussed in Section \ref{sec:conclusion}.

\section{Source separation through spectrogram morphological filtering} \label{sec:methods}
\subsection{Typical Algorithm and Oracle Method}
We investigate three promising \ac{bass} methods based on morphological filtering of the
mixture's spectrogram (defined as the squared modulus of its \ac{stft} \cite{tfbook}).
Each method aims at estimating the real-valued non-negative matrices of size $F \times T$,
which correspond to the source separation masks $M_v$, $M_h$ and $M_p$,
respectively associated to the voice, the harmonic accompaniment and the percussive part.
Thus, a typical algorithm using any BASS method, can be formulated by Algorithm \ref{alg:bass}.
%--------------- Fig1_alpha_Wiener.m
\begin{algorithm}[!ht]
 \KwData{$x$: observed mixture, $\alpha$: user parameter (\cf Fig.~\ref{fig:alpha})}
 \KwResult{$\hat{s}_i$: estimated source signals, $\hat{S}_i$: STFTs of the estimated sources}
  $X \gets \text{STFT}(x)$\\
  $(M_v, M_h, M_p) \gets \text{BASSMethod}\left(|X|^2\right)$\\
  \For{$i \in \{v,h,p\}$}
   {
     $\hat{S}_i \gets \frac{|M_i|^\alpha}{\sum_{j\in\{v,h,p\}} |M_j|^\alpha}$ X\\
     $\hat{s}_i \gets \text{invSTFT}(\hat{S}_i)$\\
   }
 \label{alg:bass}
 \caption{Typical BASS algorithm based on morphological filtering. $\text{STFT}()$ and $\text{invSTFT}()$ compute respectively the \ac{stft} and its inverse
 from a discrete-time signal.}
\end{algorithm}
%$\text{BASSMethod}()$ denotes any morphological filtering \ac{bass} method, applied on the input spectrogram.
%---------------  Fig1_
\begin{figure}[!ht]
 \subfigure[objective results]{\includegraphics[width=0.244\textwidth]{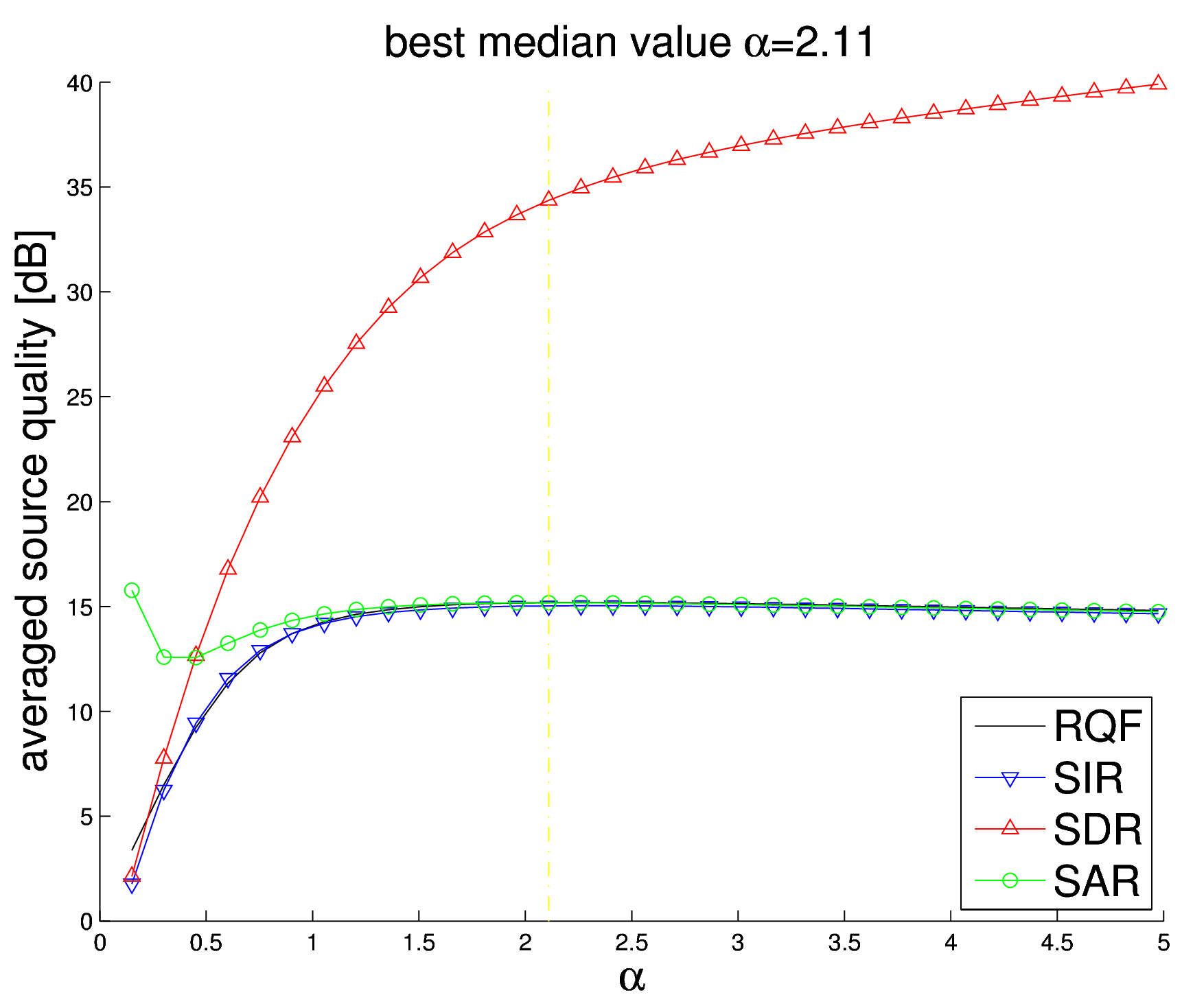}}\subfigure[objective results]{\includegraphics[width=0.244\textwidth]{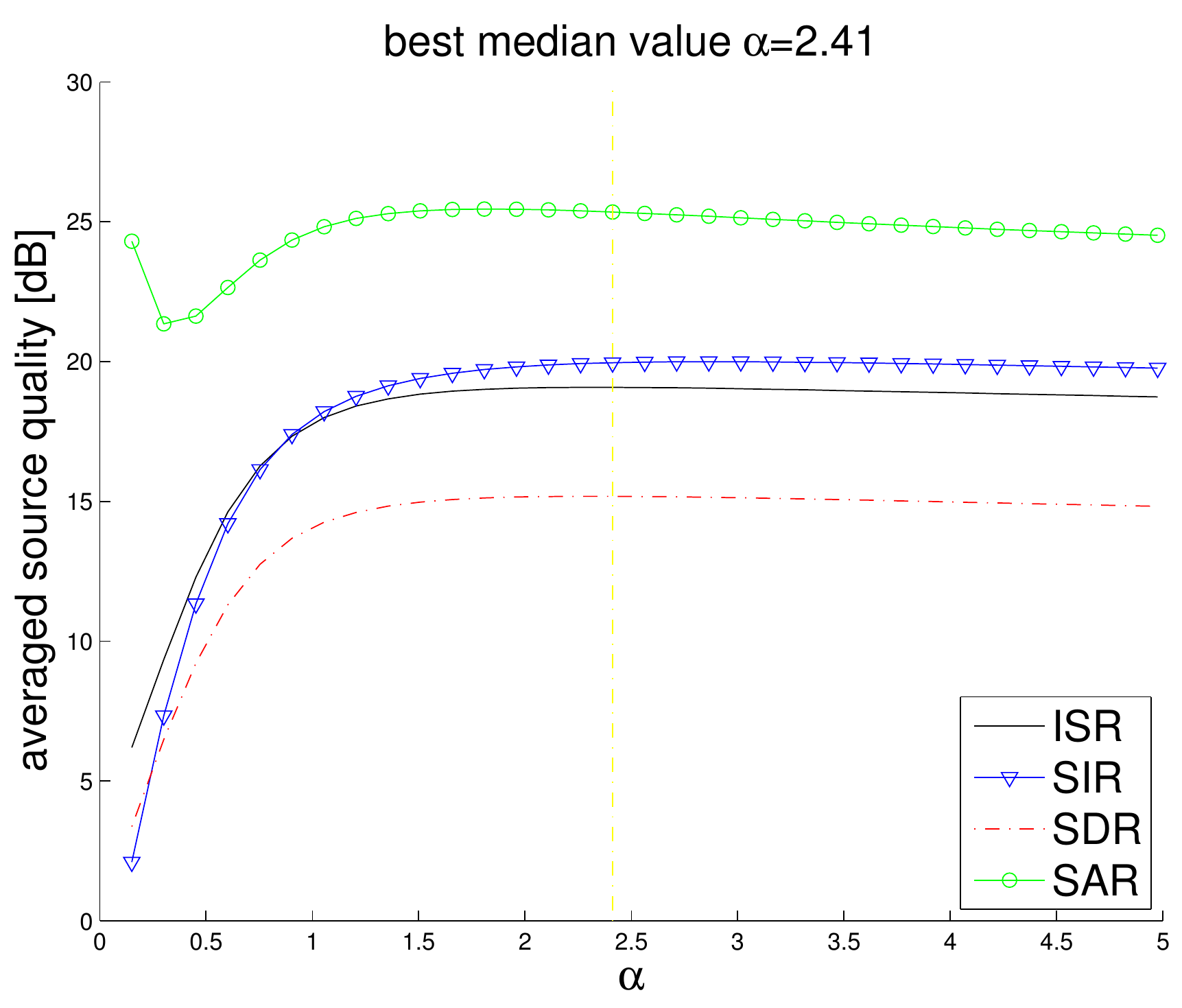}}\\
 \subfigure[perceptual results]{\includegraphics[width=0.244\textwidth]{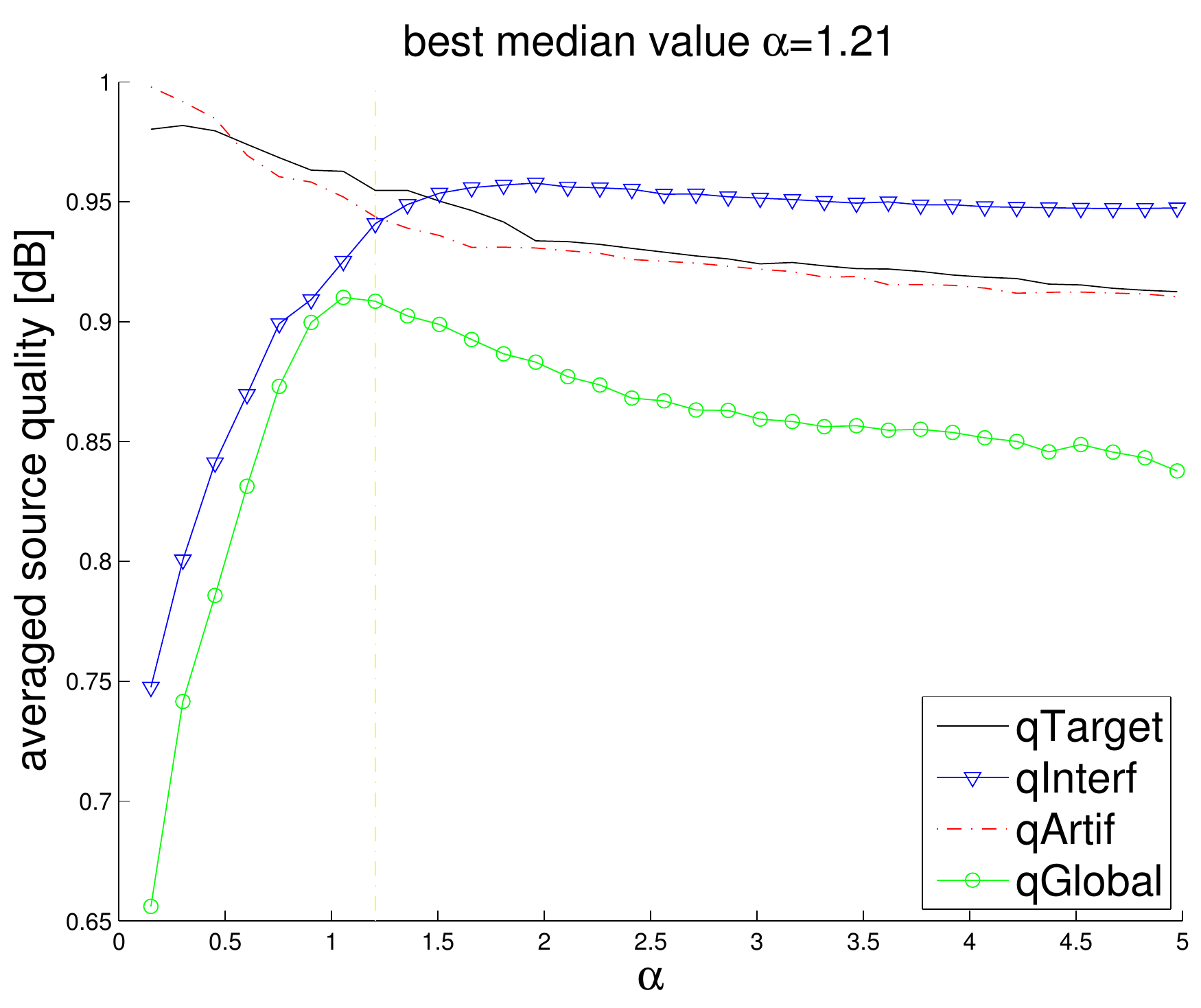}}\subfigure[perceptual results]{\includegraphics[width=0.244\textwidth]{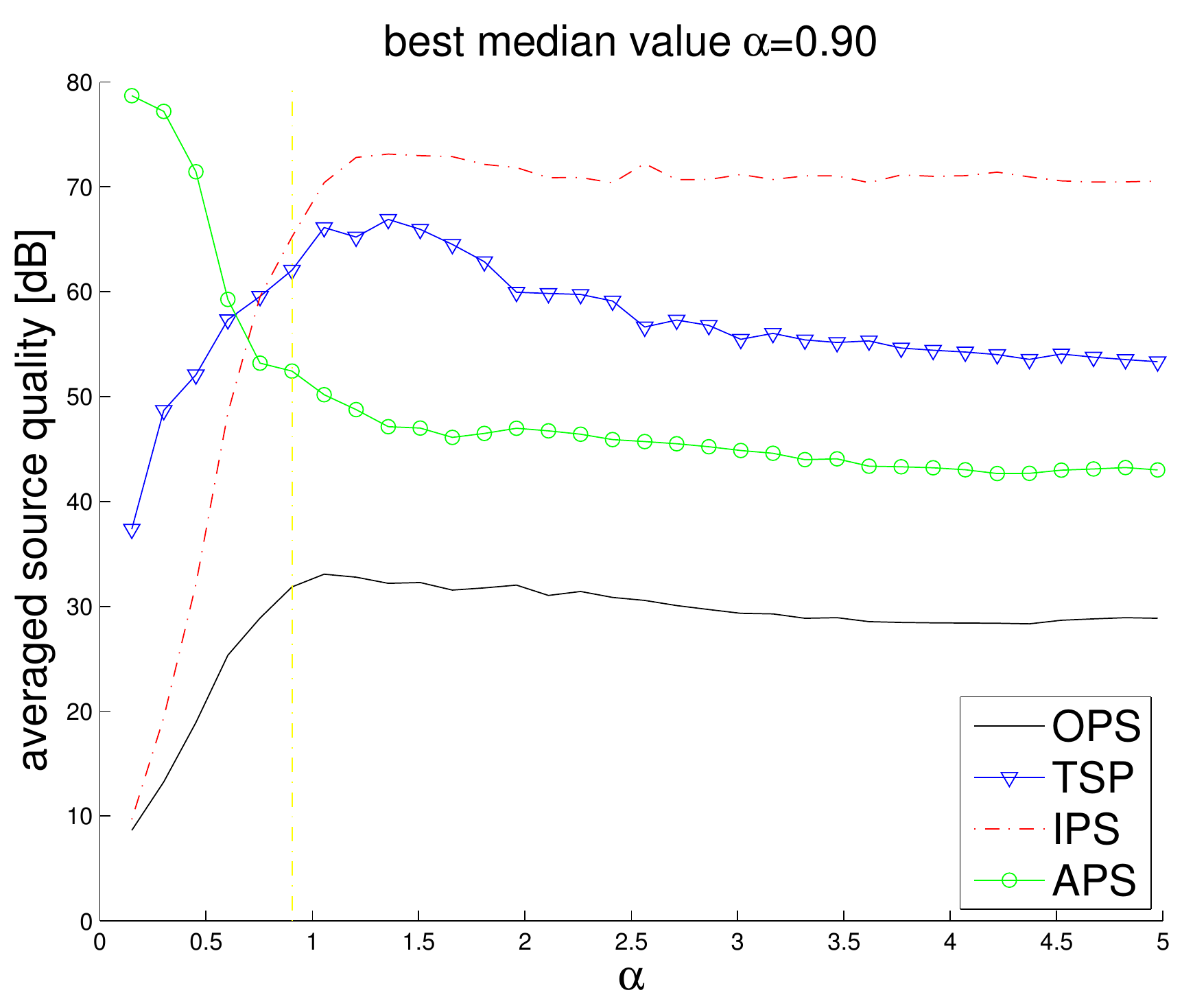}}\\
 \caption{Effect of parameter $\alpha$ in Algorithm \ref{alg:bass} on the source separation quality 
 of a musical mixture made of 3 sources.
 Measures are expressed in terms of BSS Eval v2 \cite{bsseval} (a), BSS Eval v3 \cite{emiya2011subjective} (b)-(d) which also assess the perceptual quality
 (high values are better).}
 %A formal definition of the other quality measures such as \ac{sir}, \ac{sdr} and \ac{sar} can be found in \cite{bsseval}.}
 \label{fig:alpha}
\end{figure}
%---------------
%\footnotetext[1]{BSS Eval and PEASS: \url{http://bass-db.gforge.inria.fr/bss_eval/}}
\vspace{-0.3cm}

In this algorithm, $\frac{|M_i|^\alpha}{\sum_{j\in\{v,h,p\}} |M_j|^\alpha}$ approximates
the parameterized Wiener filter \cite{fontaine2017} of the source $i$, for which an optimal value of $|M_i|^\alpha$ in the minimal
\ac{mse} sense, corresponds to the source's spectral density \cite{najim2010modeling}.
In practice, the effect of parameter $\alpha$ on the separation quality is illustrated in Fig.~\ref{fig:alpha}
which shows the results provided by Algorithm \ref{alg:bass} when applied on a mixture made of 3 audio sources (voice, keyboard/synthesizer and drums).
This experiment uses an oracle \ac{bass} method (\ie original sources are assumed known) which sets 
the source mask as the modulus of the \ac{stft} of each source such as $M_i=|S_i|$.
The highest median of the \ac{mse}-based results (\cf Fig.~\ref{fig:alpha} (a)-(b)) is reached with $\alpha \approx 2$.
Interestingly, best perceptual results are reached with $\alpha \approx 1$ (\cf Fig.~\ref{fig:alpha} (c)-(d)).
%Thus, our experiments use $\alpha=2$ to maximize objective quality measures.
A detailed description of \ac{sir}, \ac{sar} and \ac{sdr} measures can be found in \cite{bsseval,emiya2011subjective}.
The \ac{rqf} (\cf Fig.~\ref{fig:alpha} (a)) is defined as\cite{fourer2016}:
$\text{RQF}(s,\hat{s}) = 10 \log_{10}\left( \frac{\sum_n |s[n]|^2}{\sum_n |s[n]-\hat{s}[n]|^2}\right)$,
where $s$ and $\hat{s}$ stand respectively for the original source and its estimation.

\subsection{Total Variation Approach}\label{sec:jeonglee}
Blind source separation can be addressed as an optimization problem solved using a total variation regularization.
This approach has successfully been used in image processing for noise removal \cite{rudin1992nonlinear}. 
It consists in minimizing a convex auxiliary function which depends on regularization parameters
$\lambda_1$, $\lambda_2$ to control the relative importance of the smoothness of the expected masks 
$M_h$ and $M_p$ respectively over time and frequencies. This choice is justified by the harmonic or spectral stability of
$M_h$ and $M_p$, and the sparsity of $M_v$. Being a discrete-time signal $x[n]$ and its discrete \ac{stft}, $X\nm$, where 
$n=1...T$ and $m=1...F$, are the time and frequency indices such as $t=n\te$ and $\omega=2\pi\frac{m}{F\te}$,
$\te$ being the sampling period. The Jeong-Lee-14 method \cite{jeong2014vocal} minimizes the following auxiliary function:
%---------------
\begin{align}
\nonumber    J(M_v, M_h, M_p) 		&= \frac{1}{2}\sum_{n,m}(M_h[n-1,m]-M_h[n,m])^2 \\
\nonumber				&+ \frac{\lambda_1}{2} \sum_{n,m}(M_p[n,m-1]-M_p[n,m])^2\\
					&+ \lambda_2 \sum_{n,m}\left|M_v[n,m]\right|\\
\nonumber  \text{subject to:} \quad    & M_v+M_h+M_p = |X|^{2\gamma}\\
\nonumber  \text{with:}\quad 		& M_v[n,m], M_h[n,m], M_p[n,m] \geq 0.
%\nonumber				& ~~\forall (n,m) \in \mathbb{Z}^2
\end{align}
%---------------
%where $n \in [1,T]$ and $m \in [1,F]$, stand respectively for the time and the frequency indices.

Hence, solving $\frac{\partial J(M_v, M_h, M_p)}{\partial M_h}=0$ and $\frac{\partial J(M_v, M_h, M_p)}{\partial M_p}=0$, allows
to derive update rules which lead to an iterative method formulated by Algorithm \ref{alg:jeong} \cite{jeong2014vocal}.
According to the authors, the best separation results are obtained with 16 kHz-sampled signal mixtures, using 64~ms-long $\frac{3}{4}$-overlapped 
analysis frames, in combination with a 120~Hz high-pass filter applied on the mixture, and using method parameters: 
$\lambda_1=0.25$, $\lambda_2=10^{-1}\lambda_1$, $\gamma=\frac{1}{4}$ (\ie $\alpha=2$) and $N_{\text{iter}}=200$.
\vspace{-0.1cm}
%---------------
\begin{algorithm}[!ht]
 \KwData{$x$: observed mixture, $\lambda_1,\lambda_2,\gamma$: user parameters, $N_{\text{iter}}$: number of iterations}
 \KwResult{$\hat{s}_i$: estimated source signals, $\hat{S}_i$: STFTs of the estimated sources}
  $X \gets \text{STFT}(x)$\\
  $W \gets |X|^{2\gamma}$\\
  $M_h\gets 0$, $M_p\gets 0$\\
  \For{$it \gets 1$~to~$N_{\text{iter}}$}
  {
    $M_h{\scriptstyle [n,m]} \gets \min\!\left(\!\frac{M_h{\scriptstyle[n+1,m]}+M_h{\scriptstyle [n-1,m]}}{2} + \frac{\lambda_1}{2},\right.$\\
    \hfill $\left. W{\scriptstyle [n,m]}-M_p{\scriptstyle [n,m]} \right)$\\
    $M_p{\scriptstyle[n,m]} \gets \min\!\left(\!\frac{M_p{\scriptstyle[n,m+1]}+M_p{\scriptstyle [n,m-1]}}{2} + \frac{\lambda_1}{2\lambda_2},\right.$\\ 
    \hfill $\left. W{\scriptstyle [n,m]}-M_h{\scriptstyle [n,m]} \right)$\\
  }
  $M_v \gets W - (M_h+M_p)$\\
  \For{$i \in \{v,h,p\}$}
  {
    $\hat{S}_i \gets \frac{|M_i|^{\frac{1}{2\gamma}}}{\sum_i |M_i|^{\frac{1}{2\gamma}}}$ X\\
    $\hat{s}_i \gets \text{invSTFT}(\hat{S}_i)$\\
  }
\caption{Jeong-Lee-14's \ac{bass} algorithm.}
\label{alg:jeong}
\end{algorithm}
%---------------
\vspace{-0.5cm}
\subsection{Robust Principal Component Analysis}\label{sec:rpca}

In a musical mixture, the background accompaniment is often repetitive while the main melody
played by the singing voice contains harmonic and frequency modulated components with 
a non-redundant structure. This property allows a decomposition of the mixture spectrogram 
$W=|X|^2$ into two distinct matrices where the background accompaniment spectrogram is associated to a low rank matrix,
and the foreground singing voice is associated to a sparse matrix (\ie where most of the elements are zeros or close to zero).
Thus, a solution inspired from the image processing methods is provided by \ac{rpca} \cite{candes2011robust} 
which decomposes a non-negative matrix $W$ into a sum of two matrices $M_{hp}$ and $M_v$, through
an optimization process.
It can be formulated as the minimization of the following auxiliary function expressed as: %with $L$ is a low rank matrix, and $S$ is a sparse matrix .
% ------------------------ % \text{minimize~} 
%\vspace{-0.1cm}
\begin{align}
	    J(M_{hp},M_v)	&= ||M_{hp}||_* + \lambda ||M_v||_1 \label{eq:rpca01} \\
 \nonumber \text{subject to:~}	& W = M_{hp} + M_v
\end{align}
% ------------------------
with $||M_{hp}||_* = \sum_k \sigma_k(M_{hp})$ the nuclear norm of matrix $M_{hp}$, $\sigma_k$ 
being its $k$-th singular value, and $||M_v||_1=\sum_{n,m} |M_v[n,m]|$ being the $l_1$-norm of the matrix $M_v$.
Here, $\lambda$ denotes a damping parameter which should be optimally chosen as $\lambda=\frac{1}{\sqrt{\max(T,F)}}$ \cite{candes2011robust,huang2012singing}.
% When $X$ is a mixture spectrogram, this technique allows to separate the background accompaniment from the foreground
% singing voice. As the accompaniment is often repetitive, its spectrogram is associated to the low rank matrix, while the
% sparse matrix corresponds to the singing voice (or the main melody).
Eq. \eqref{eq:rpca01} is then solved by the augmented Lagrangian method which leads to the following new auxiliary function (adding new variable $Y$):
% ------------------------
\vspace{-0.1cm}
\begin{multline}
 J(M_{hp},M_v,Y) = ||M_{hp}||_* + \lambda ||M_v||_1 +\\ \langle Y,W-M_{hp}-M_v \rangle + \frac{\mu}{2}||W-M_{hp}-M_v||^2_F
 \label{eq:augrpca}
\end{multline}
% ------------------------
where $\langle a,b\rangle =a^Tb$, and $\mu$ is a Lagrangian multiplier. % and $||X||_*=\sum_{i=1}^{\min(T,F)}\sigma_i(X)$ denotes the nuclear norm of $X$ computed by the sum of its singular values.
Thus, Eq. \eqref{eq:augrpca} is efficiently minimized through the Principal Component Pursuit algorithm \cite{pcpursuit2009} formulated by
Algorithm \ref{alg:pcp}. Our empirical experiments on real-word audio signals show that $\mu=10\lambda$ and $N_\text{iter}=1000$ provide satisfying results.
% ------------------------
\begin{algorithm}[!ht]
 \KwData{$W$: spectrogram of the mixture, $\lambda, \mu$: damping parameters, $N_{\text{iter}}$: number of iterations}
 \KwResult{$L=M_{hp}$, $S=M_v$: separation masks for the voice (v) and the music accompaniment (hp)}
  $S \gets 0$, $Y \gets 0$\\
  \For{$it \gets 1$~to~$N_{\text{iter}}$}
  {
   $L \gets \argmin_L J(L,S,Y)$\\
   $S \gets \argmin_S J(L,S,Y)$\\
   $Y \gets Y + \mu (W-L-S)$
  }
  \caption{Principal Component Pursuit by alternating directions algorithm \cite{pcpursuit2009}.}
  \label{alg:pcp}
\end{algorithm}
% ------------------------
\vspace{-0.2cm}

For the sake of computation efficiency, it can be shown that the update rules in Algorithm \ref{alg:pcp} 
can be computed as \cite{candes2011robust}:
% ------------------------
\begin{align}
 \argmin_L J(L,S,Y)		 	&= \mathcal{S}_{\lambda\mu^{-1}}(W-L+ \mu^{-1}Y)\\
 \nonumber\text{with~}		 	& \mathcal{S}_\tau(x) = \text{sign}(x)\max(|x|-\tau,0)\\
\nonumber\\
 \argmin_S J(L,S,Y)	 		&= \mathcal{D}_{\mu^{-1}}(W-S+ \mu^{-1}Y)\\
 \nonumber\text{with~}			& \mathcal{D}_\tau(X) = U\mathcal{S}_\tau(\Sigma)V^*
\end{align}
% ------------------------
where $X = U \Sigma V^*$ is the singular value decomposition of matrix $X$ and $V^*$
denotes the conjugate transpose of matrix $V$ (\ie $V$ is the matrix where each column is a right-singular vector).
Finally, each source signal is recovered using the estimated separation masks $M_v$ (equal to
the sparse matrix $S$) and $M_{hp}$ (equal to the low-rank matrix $L$), through the parameterized Wiener 
filter applied on the \ac{stft} of the mixture as in Algorithm \ref{alg:bass}.

\vspace{-0.3cm}
\subsection{Kernel Additive Modeling} \label{sec:kam}
The \ac{kam} approach \cite{kim2015,liutkus2014} is inspired from the locally weighted regression theory \cite{cleveland1988locally}.
The main idea assumes that the spectrogram of a source is locally regular. In other words, it means that the vicinity of 
each time-frequency point $(t,\omega)$ in a source's spectrogram can be predicted. 
Thus, the \ac{kam} framework allows to model source-specific assumptions such as the harmonicity of a source 
(characterized by horizontal lines in the spectrogram), percussive sounds (characterized by vertical lines in the spectrogram)
or repetitive sounds (characterized by recurrent shapes spaced by a time period in the spectrogram).
A \ac{kam}-based source separation method can be implemented according to Algorithm \ref{alg:kam_bss} using 
the desired source-specific kernels $\mathcal{K}^b_i$ corresponding to binary matrices of size $h \times w$ as illustrated in Fig.~\ref{fig:kam}.\\
% ------------------------
\begin{algorithm}[!ht]
 \KwData{$X$: mixture STFT, $\mathcal{K}^b_i$: kernel of each source $i\leq I$, $\alpha$: user parameter, $N_{\text{iter}}$: number of iterations}
 \KwResult{$\hat{s}_i$: estimated source signals, $\hat{S}_i$: STFTs of the estimated sources}
  $\hat{S}_i \gets \frac{X}{I}$, $~\forall i \in [1,I]$\\
  \For{$it \gets$ $1$ to $N_{\text{iter}}$ }{
  \For{$n \gets$ $1$ to $T$ \textbf{and} $m \gets$ $1$ to $F$}
   {
    \For{$i \gets$ $1$ to $I$}
    {
      $M_i \gets \text{median}\left|\hat{S}_i[n+c'-\frac{w-1}{2},m+l'-\frac{h-1}{2}]\right|$, $\left\{(c',l') : \mathcal{K}^b_i(c',l') = 1\right\}$
    }
    $\hat{S}_i\nm \gets \frac{|M_i|^\alpha}{\sum_{j=1}^{I} |M_j|^\alpha} X\nm$, $\forall i \in [1,I]$
   }
  }
  $\hat{s}_i \gets \text{invSTFT}\left(\hat{S}_i\right)$, $\forall i \in [1,I]$\\ %\exp(j \cdot \arg(X)) 
  %\vspace{0.2cm}
 \caption{KAM-based source separation algorithm.}  %$j^2=-1$., $\arg(X)$ being the phase of complex matrix $X$ and
 \label{alg:kam_bss}
\end{algorithm}
% ------------------------
\vspace{-0.2cm}
\subsubsection{How to choose a Kernel for source separation?}
\begin{figure}[!ht]
\centering\includegraphics[width=0.49\textwidth]{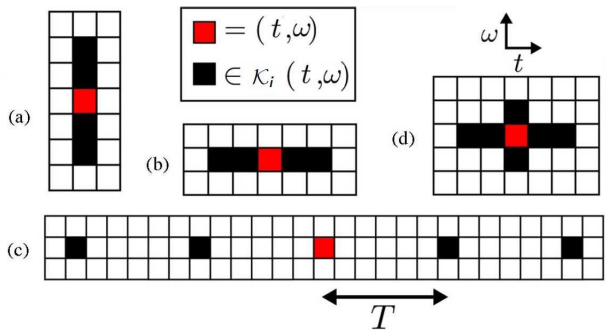}
\caption{Illustration of several possible kernels \cite{liutkus2014}, (a) for percussive sources, (b) for harmonic sources, (c) for repetitive elements and (d) for smoothly varying sources (e.g. vocal).}
 \label{fig:kam}
\end{figure}
% ------------------------
\begin{figure}[!ht]
\centering\includegraphics[width=0.49\textwidth]{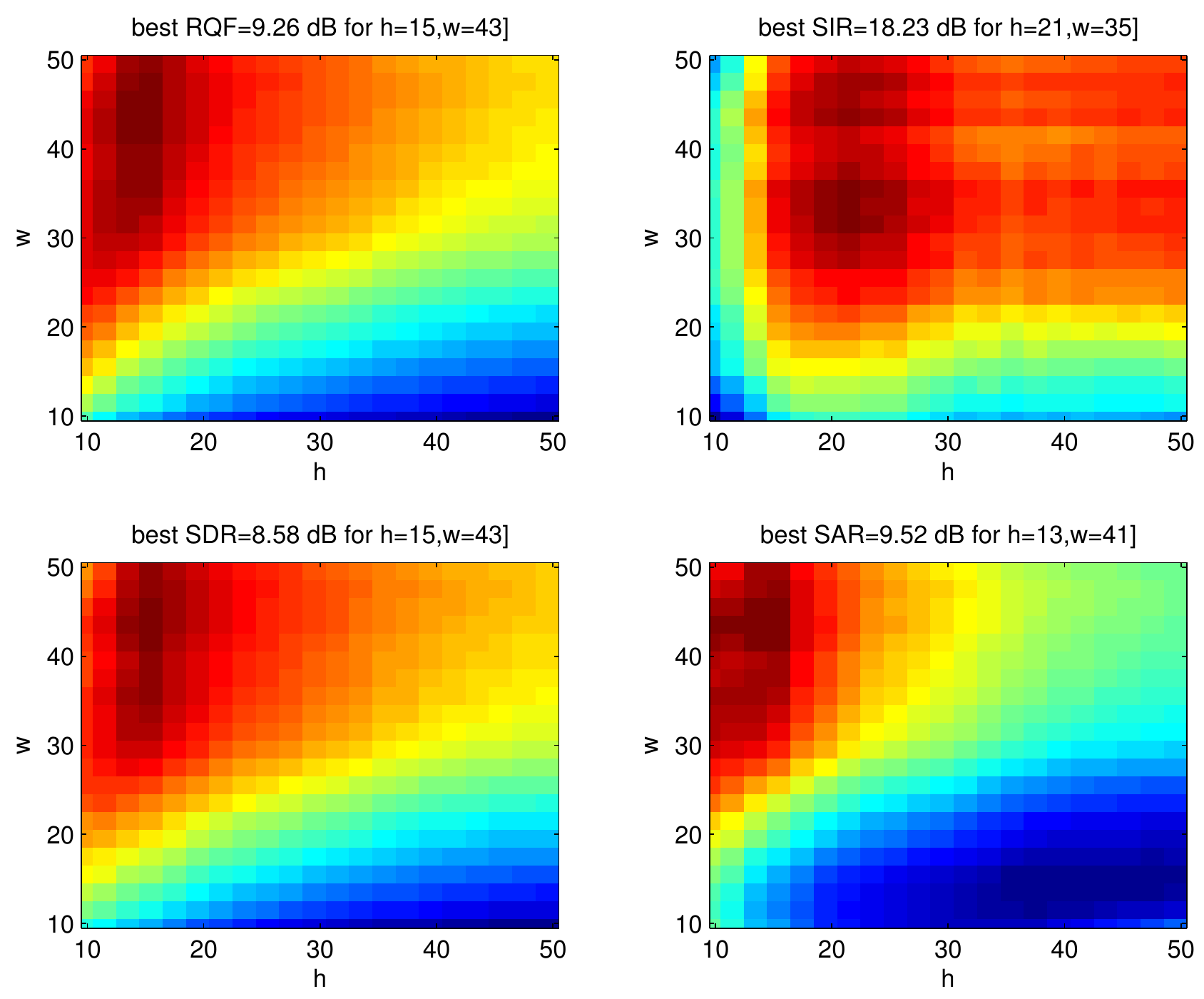}
 \caption{Comparison of the separation quality, measured in terms of RQF, SIR, SDR, SAR \cite{bsseval}, as a function of 
  $h$ and $w$, the dimensions of the separation kernels. We considered a musical piece made of 2 sources (voice/drums). 
 A darker red color corresponds to a better separation quality.}
 \label{fig:kamwh}
\end{figure}
% ------------------------

As a kernel aims at modeling the vicinity at each point of a time-frequency representation, several typical 
kernels can be extracted from the literature as presented in Fig.~\ref{fig:kam}.
\ac{hpss} methods using median filtering \cite{fitzgerald2010harmonic, fitzgerald2014harmonic} can use: \textbf{(a)+(b)}.
Algorithms such as the REPET algorithm \cite{rafii2013repeating,liutku2012}, which can separate vocal from accompaniment uses: \textbf{(c)+(d)}.
These methods use the repetition rate denoted $\textbf{T}$ in Fig.~\ref{fig:kam}, corresponding to the music tempo.
For a musical piece $\textbf{T}$ can be constant such as proposed in \cite{rafii2013repeating} or time-varying (adaptive) as in \cite{fitzgerald2014harmonic}.
% %------------------------

Another question is how to choose the size of a kernel in order to optimize the separation quality?
An empirical answer provided by grid search is illustrated in Fig.~\ref{fig:kamwh} which shows the best choice for $h$ and $w$,
to maximize the separation quality measures (RQF, SIR, SDR, SAR). For this experiment the \ac{stft}
of a signal sampled at $\fs=22.05$ kHz is computed using a Hann window of length $N = 2048$ samples ($\approx$92 ms) and an overlap ratio between adjacent frames equal to $\frac{3}{4}$. 
The separation is obtained using two distinct kernels (\cf Fig.~\ref{fig:kam} \textbf{(a)+(b)}), to provide 2 sources from a mixture made of a singing voice signal and drums.
In this experiment, the best \ac{sir} equal to 18.23 dB is obtained with $h=21$ and $w=35$.
This is an excellent separation quality in comparison with the oracle \ac{bass} method used in Fig.~\ref{fig:alpha}.
\ac{rqf}, \ac{sdr} and \ac{sar} related to signal quality, are also satisfying but not optimal. % for $h=21$ and $w=35$. %which are more related to the source quality than separation efficiency,
\vspace{0.2cm}

\subsubsection{Towards a training method for supervised KAM-based source separation}
\begin{figure}[!ht]
 \centering\includegraphics[width=0.49\textwidth]{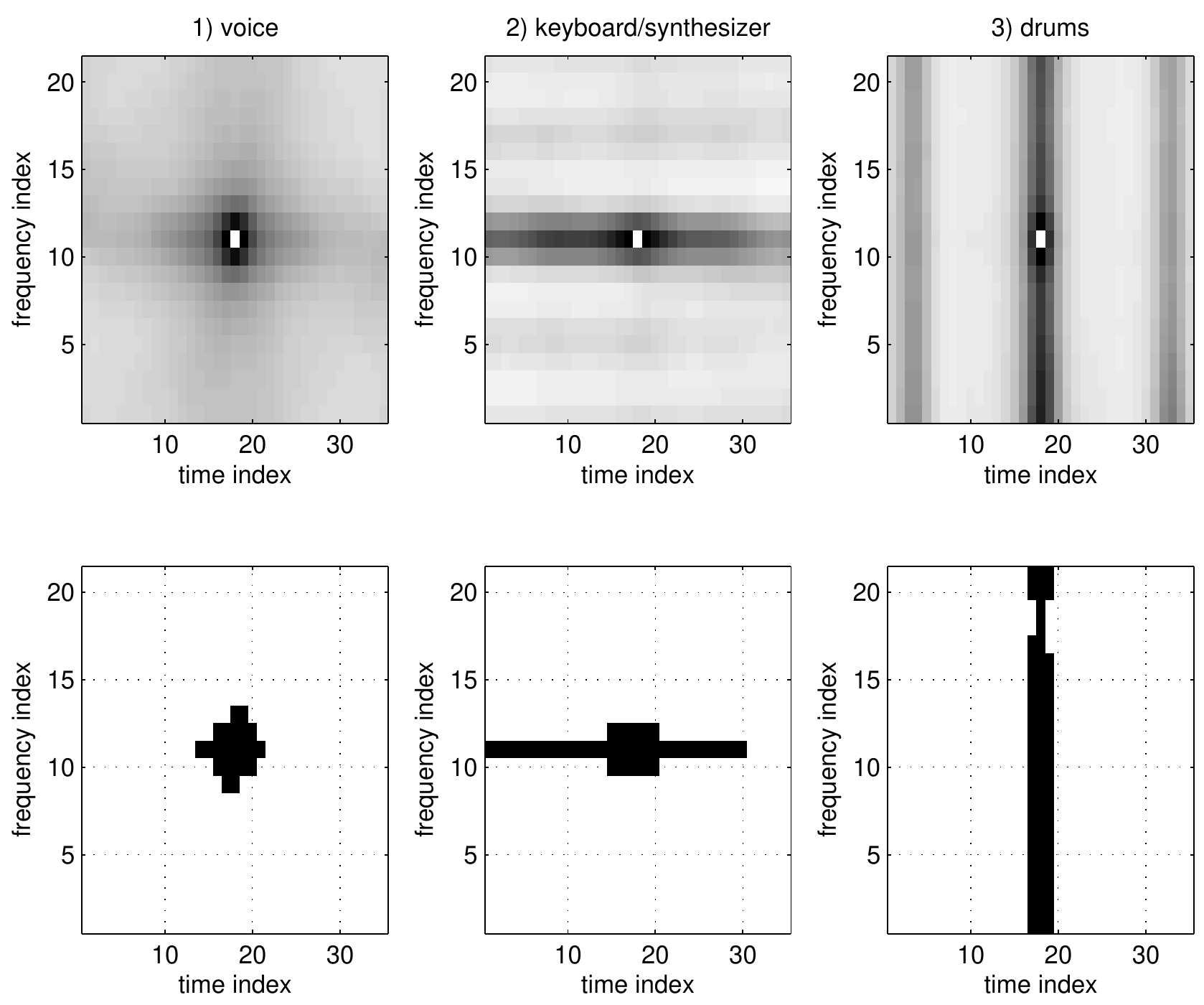}
 \caption{Kernels provided by Algorithm \ref{alg:train_kam} with $\Gamma=0.54$, $h=21$, $w=35$, applied on a mixture of 3 sources: 1) singing voice, 2) keyboard/synthesizer and 3) drums. 
 The first row corresponds to $\mathcal{K}_i$ and the second one to $\mathcal{K}^b_i$.}
 \label{fig:train_kam}
\end{figure}

To the best of our knowledges, no dedicated method exists to automatically define the best source-specific kernel
to use through a KAM-based BASS method. Hence, a classical approach consists of an empirical choice
of a predefined typical kernel and of its size.
To this end, we propose a new method depicted by Algorithm \ref{alg:train_kam},
which provides a source-specific kernel $\mathcal{K}^b_i \in \{0,1\}^{h\times w}$ associated to the source $i$.
The main idea consists in modeling the vicinity of each time-frequency point
through an averaged neighborhood map obtained after visiting each coordinate of a source spectrogram. 
%This averaged map is obtained by weighting each vicinity map by the energy present in the current visited time-frequency point.
The resulting kernel denoted $\mathcal{K}_i\in\mathbb{R}^{h\times w}$ is then binarized in order to be directly used 
by the KAM method, through a user-defined threshold $\Gamma$ such as:
% ------------------------
\vspace{-0.1cm}
\begin{equation}
 \mathcal{K}^b_i[c,l] = \begin{cases} 1 & \text{if ~} \mathcal{K}_i[c,l] > \Gamma\\ 0 & \text{otherwise}\end{cases}.
\end{equation}
% ------------------------
Our new method based on customized kernels (KAM-CUST) is applied on musical signals in Fig.~\ref{fig:train_kam}.
The results clearly illustrate the different trained source-specific kernels between singing voice, keyboard/synthesizer and drums as in Fig. \ref{fig:kam}.
%------------------------
\vspace{-0.4cm}
%------------------------
\begin{algorithm}[!ht]
 \KwData{$S_i$: a source STFT}
 \KwResult{$\mathcal{K}_i \in \mathbb{R}^{h\times w}$, $h$ and $w$ being odd integers.}
 %$\mathcal{K}_s[c,l] \gets 0$, $\forall c \in [1,w]$ and $\forall l \in [1,h]$.
 $K_j[c,l]\gets 0$, $\forall c \in [1,w]$, $\forall l \in [1,h]$, and $\forall j \in [1, TF]$\\
 $p_j \gets 0, \forall j \in [1, TF]$\\
 $j \gets 1$\\
 \For{$n \gets$ $1$ to $T$ \textbf{and} $m \gets$ $1$ to $F$}
 {
   $K_j \gets \left|S_j[n-\frac{c-1}{2}:n+\frac{c-1}{2},m-\frac{h-1}{2}:m+\frac{h-1}{2}]\right|$\\
   $K_j \gets \frac{K_j}{||K_j||}$\\
   $p_j \gets |S_i\nm|^2$\\
   $j \gets j+1$
 }
 \For{$c \gets$ $1$ to $w$ \textbf{and} $l \gets$ $1$ to $h$}
 {
  $\mathcal{K}_i[c,l] \gets \frac{\sum_{j=1}^{TF} K_j[c,l] p_j}{\sum_{j=1}^{TF} p_j}$
 }
 \caption{KAM training algorithm$^1$.} %$(1:n)$ denote the set of integer in $[1,n]$.
 \label{alg:train_kam}
\end{algorithm}
% %------------------------
%% Fig_myKAM.m & Fig_myKAM_Niter
\begin{figure*}[!ht]
 \subfigure[voice]{\includegraphics[width=.33\textwidth]{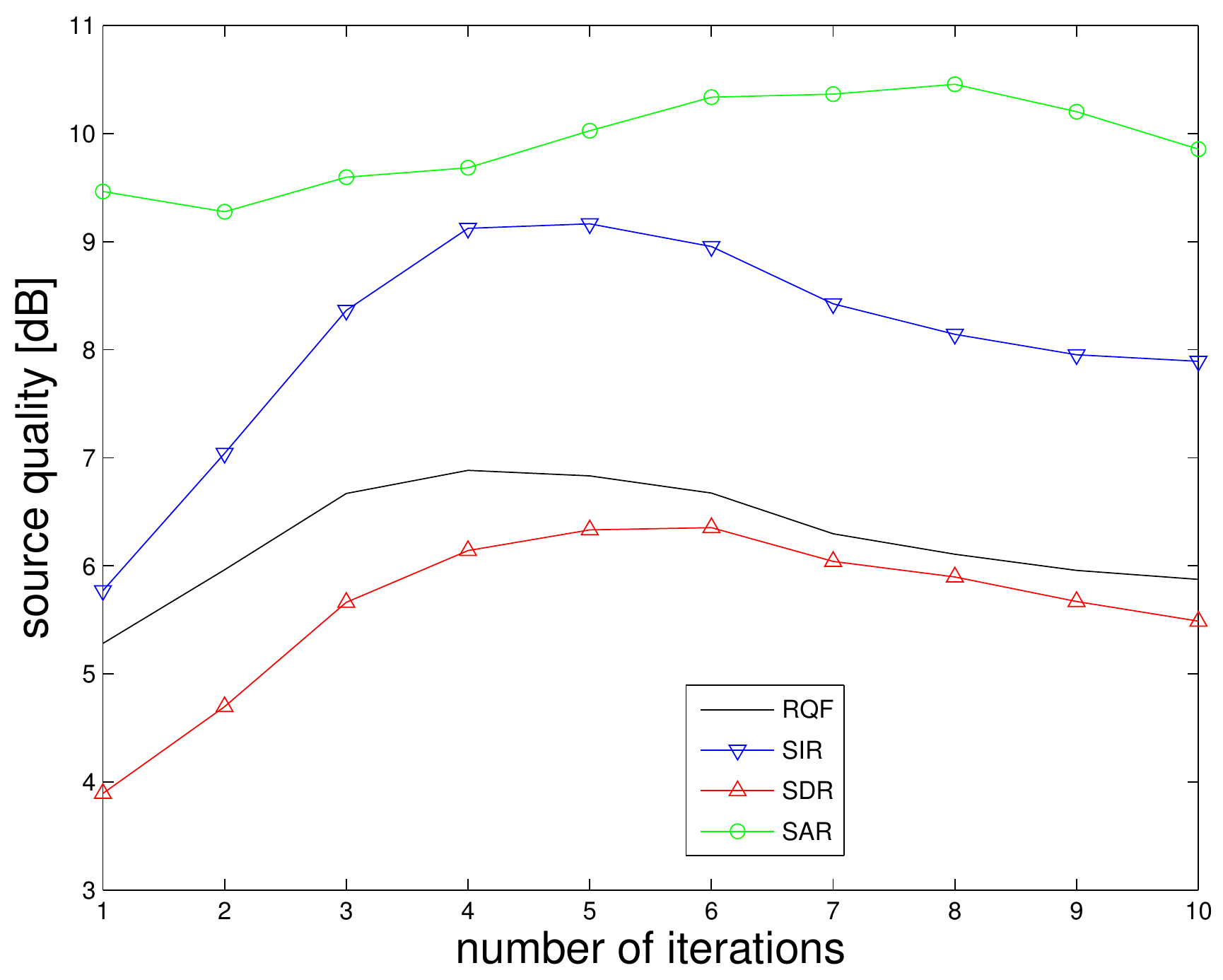}}
 \subfigure[keyboard/synthesizer]{\includegraphics[width=.33\textwidth]{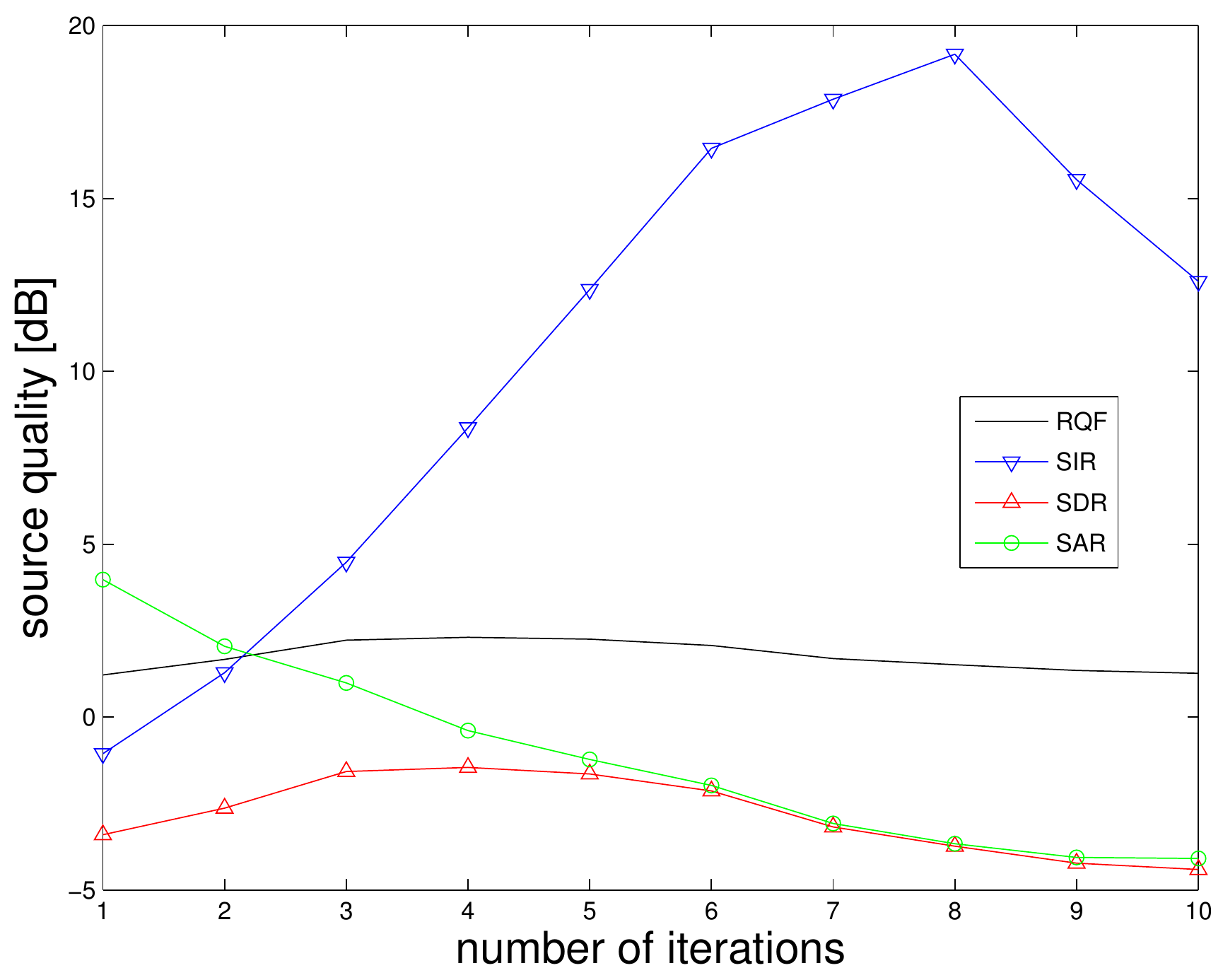}}
 \subfigure[drums]{\includegraphics[width=.33\textwidth]{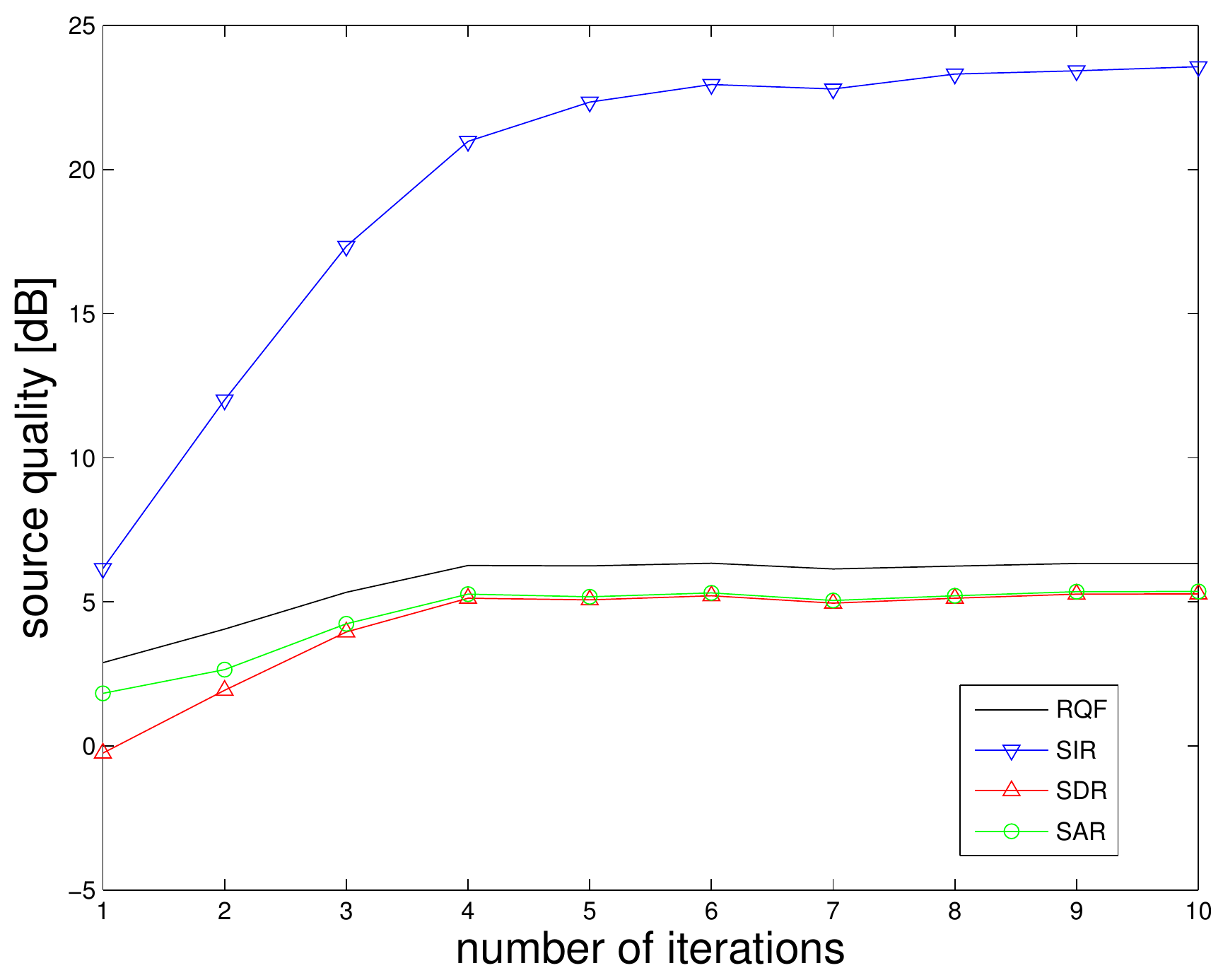}}
 \caption{Separation quality using trained kernels on a mixture made of 3 sources as a function of the number of iterations $N_{\text{iter}}$.}
 \label{fig:kamiter}
\end{figure*}
% ------------------------
 %\vspace{0.5cm}
%%--------------------------------------------------------------------------------
\begin{table}[!ht]
\caption{Separation of a mixture made of 3 sources using different KAM configurations.}
\!\subfigure[new proposed (KAM-CUST) semi-blind approach using the 3 trained kernels in Fig.~\ref{fig:train_kam} ($h=21$, $w=35$, $N_{\text{iter}}=4$, $\alpha=2$)]{
 \resizebox{0.49\textwidth}{!}{
\begin{tabular}{|c|c|c|c|c|}
\hline
 Source						& RQF (dB)			& SIR (dB)			& SDR (dB)			& SAR (dB)\\
 \hline
 voice						& \textbf{6.88}			& 9.12				& \textbf{6.14}			& \textbf{9.68}\\
 \hline
 keyboard					& \textbf{2.31}			& \textbf{8.36}			& -1.45				& -0.38\\
 \hline
 drums						& \textbf{6.26}			& \textbf{20.98}		& \textbf{5.12}			& \textbf{5.27}\\
  \hline
 \end{tabular}}}
\!\subfigure[KAM method using REPET kernels \cite{liutku2012,liutkus2015} combined with \ac{hpss}\cite{fitzgerald2010harmonic}.]{
\resizebox{0.49\textwidth}{!}{
\begin{tabular}{|c|c|c|c|c|}
\hline
 Source						& RQF (dB)			& SIR (dB)			& SDR (dB)			& SAR (dB)\\
 \hline
 voice						& 3.16				& \textbf{10.33}		& 0.30				& -1.14\\
 \hline
 keyboard					& 0.89				& 4.67				& \textbf{-1.4}			& \textbf{1.10}\\
 \hline
 drums						& -3.20				& 3.01				& -3.36				& -0.47\\
  \hline
 \end{tabular}}}
 \!\subfigure[KAM method using REPET kernels \cite{liutku2012,liutkus2015}, without HPSS]{
 \resizebox{0.49\textwidth}{!}{
\begin{tabular}{|c|c|c|c|c|}
\hline
 Source						& RQF (dB)			& SIR (dB)			& SDR (dB)			& SAR (dB)\\
 \hline
 voice.						& 4.76				& 8.06				& 3.33				& 5.74\\
 \hline
 keyb.+drums					& 1.09				& 4.04				& -2.94				& -0.52\\
 \hline
 \end{tabular}}}
 \label{tab:kam_bss}
\end{table}
%%--------------------------------------------------------------------------------
 \footnotetext[1]{$A[a:b,c:d]$ denotes the submatrix of $A$ such as $\left(A[i,j]\right)_{i\in[a,b],j\in[c,d]}$}
% %------------------------

To show the efficiency of this training method, we apply Algorithm \ref{alg:train_kam}
on each isolated component of the same mixture as before made of 3 sources (voice, keyboard/synthesizer and drums)
sampled at $\fs = 22.05$ kHz. The resulting trained kernels displayed in Fig.~\ref{fig:train_kam}
are then used in combination with Algorithm \ref{alg:kam_bss} for KAM-based \ac{bass}.
In this experiment, we compare the separation results obtained by our proposal (KAM-CUST) with $h=21$, $w=35$, $N_{\text{iter}}=4$, $\alpha=2$ (\cf Table \ref{tab:kam_bss} (a)),
with the results provided by the KAM-REPET algorithm as implemented by Liutkus \cite{liutku2012,liutkus2015} (\cf Table \ref{tab:kam_bss} (c))
and when KAM-REPET is combined with the \ac{hpss} method \cite{fitzgerald2010harmonic} in order to obtain 3 sources (\cf Table \ref{tab:kam_bss} (b)).

The results show that the KAM method combined with trained kernels can significantly outperforms others state-of-the-art methods, particularity in terms of \ac{rqf}, \ac{sir}.
Our method also obtains acceptable \ac{sdr} and \ac{sar} (above 5 dB except for the keyboard recovered signal).
On the other side, the best \ac{sir} result (characterized by a better source isolation) for the extracted singing voice signal,
is provided by the combination of the REPET with the \ac{hpss} method. However, this approach obtains a poor \ac{sdr} and \ac{sar} results
and a lower \ac{rqf} than using our proposal. Hence, low \ac{sdr} and low \ac{sar} correspond to a poor perceptual audio signal quality where the original
signal is altered by undesired artifacts (\ie undesired sound effects and additive noise).

The impact of the number of iterations $N_{\text{iter}}$ using KAM-CUST is investigated in Fig.~\ref{fig:kamiter} which shows that the best \ac{rqf} for 
the extracted singing voice can be reached for $N_{\text{iter}}=4$.
A higher value of $N_{\text{iter}}$ increases the computation time and can improve the \ac{sir} of the accompaniment (which corresponds to a better separation), 
however it can also add more distortion and artifacts as shown by the \ac{sdr} and \ac{sar} curves which decrease when $N_{\text{iter}}>4$ for the resulting sources.

\section{Singing voice detection}\label{sec:voice}

\begin{figure}[!ht]
\hspace{-0.1cm}\resizebox{0.5\textwidth}{!}{\includegraphics{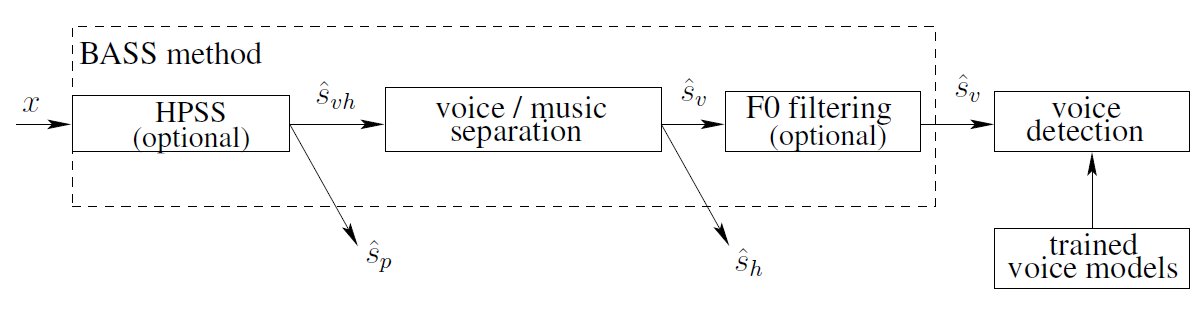}}
\caption{Proposed framework for music source separation and singing voice detection from a polyphonic mixture $x$.
HPSS \cite{fitzgerald2010harmonic} is only used separately when this capability
is not included with the \ac{bass} method (\ie KAM-REPET and RPCA).
Trained voice models are only used by the supervised approaches.}
\label{fig:baseline}
\end{figure}

In this section, we propose several approaches to detect at each time instant
if a singing voice is active into a polyphonic mixture signal.
The proposed framework illustrated by Fig.~\ref{fig:baseline} 
uses source separation as a preliminary step before applying a 
singing voice detection. We choose to investigate both the unsupervised approach and
the supervised approach which uses trained voice models to help
the recognition of signal segments containing voice.
% --------------------
%\vspace{0.2cm}
% --------------------
\vspace{-0.5cm}
\subsection{Unsupervised Singing Voice Detection}\label{sec:voice_unsupervised}
In the unsupervised approach, we do not train specific model for singing voice detection.
We only compute a \ac{vtmr} on the estimated signals provided by the \ac{bass} methods$^2$.
\footnotetext[2]{Note that in the case of KAM-CUST, the separation model is trained.}
The \ac{vtmr} is a saliency function which is computed on non-silent frames.
Thus, two user-defined thresholds are used respectively for silence detection $\Gamma_{\text{s}}$ 
and for voice detection $\Gamma_{v}$.
The voice detection process can thus be described as follows for an input signal mixture $x$.

\begin{figure}[!ht]
 \includegraphics[width=0.48\textwidth]{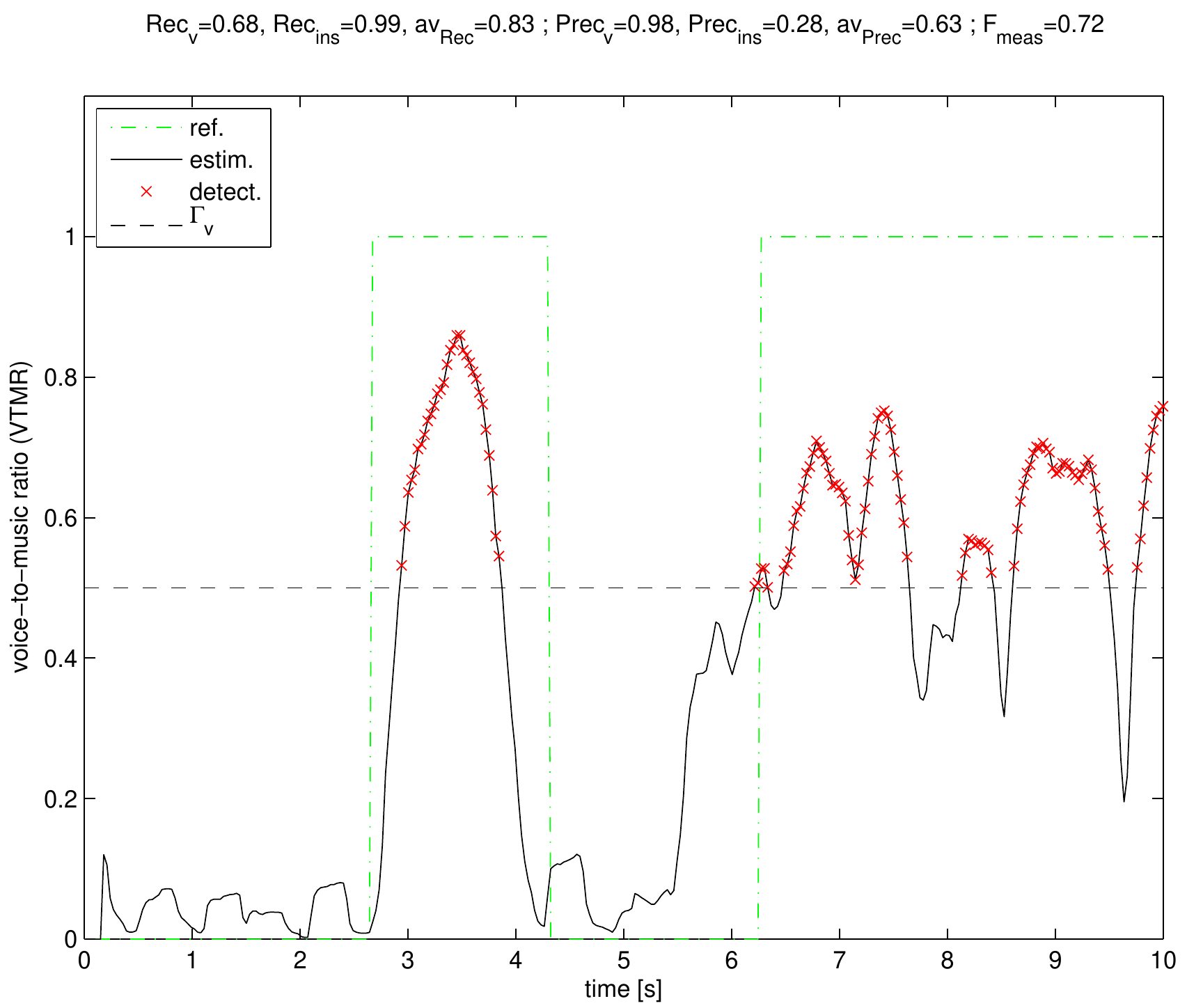}
 \caption{Unsupervised voice detection using KAM-REPET for \ac{bass}, applied on the annotated track \textit{MusicDelta Punk} taken from MedleyDB ($\Gamma_{v}=0.5$).}
 \label{fig:VDdemo}
\end{figure}

\begin{enumerate}
 \item Computation of $\hat{s}_v$ and $\hat{s}_{hp}=\hat{s}_h+\hat{s}_p$, respect to $x=\hat{s}_v+\hat{s}_{hp}$, using one of the previously proposed \ac{bass} method in Section \ref{sec:methods}.
 \item Application of a band-pass filter on $\hat{s}_v$ to allow frequencies in range $[120, 3000]$ Hz (adapted to a singing voice bandwidth).
 \item Computation of the \ac{vtmr} on each signal frame of length $N_v$ by step $\Delta_n$, centered on sample $n$, as: 
 \begin{align}
  \nonumber	E[n]	&= \displaystyle\sum_{k=n-\frac{N_v}{2}}^{n+\frac{N_v}{2}} |x[k]|^2 \label{eq:vtmr} \\
  \text{VTMR}[n] 	&= \begin{cases} \frac{\displaystyle\sum_{k=n-\frac{N_v}{2}}^{n+\frac{N_v}{2}} |\hat{s}_v[k]|^2}{E[n]},  & \text{if~} E[n] > \Gamma_{\text{s}} \\ 0 & \text{otherwise} \end{cases}
 \end{align}
 \item The decision to consider if the frame center at time index $n$ contains a singing voice is taken when $\text{VTMR}[n] > \Gamma_{v}$, with $\Gamma_{v} \in [0,1]$. Otherwise,
 an instrumental or a silent frame is considered.
\end{enumerate}
%
%\vspace{0.3cm}
Hence, in our method we assume that despite errors for estimating the voice signal $\hat{s}_v$, its corresponding energy
computed on a frame provides sufficiently relevant information to detect the presence of a singing voice in the analyzed mixture.
According to this assumption, the selected threshold $\Gamma_v$ related to \ac{vtmr} should be chosen close to 0.5.
A lower value is however less restrictive but can provide more false positive results.
About the silent detection threshold $\Gamma_s$, a low value above zero should be chosen to increase robustness to estimation errors
and to avoid a division by zero in Eq. \eqref{eq:vtmr}. Hopefully, this parameter has shown a weak importance on the voice detection results
when it is chosen sufficiently small (\eg $\Gamma_s=10^{-4}$).
An illustration of the proposed framework using the KAM-REPET \ac{bass} method is presented in Fig.~\ref{fig:VDdemo} which displays the VTMR (plotted in black) 
computed for the musical excerpt \textit{MusicDelta Punk} taken from the MedleyDB dataset \cite{MedleyDB}. 
The annotation (also called ref.) is plotted in green and the frames which are detected as containing singing voice correspond
to red crosses. In this short excerpt (\cf Fig.~\ref{fig:VDdemo}), results are excellent since the average recall is 0.83,
the average precision is 0.63 and the F-measure is equal to 0.72. Further explanations about these evaluation metrics
are provided in Section \ref{sec:singingvoice}.
%This point is more discussed with our numerical results.
%
%%% demo_VD
\vspace{-0.5cm}
\subsection{HPSS and $F_0$ Filtering}
In the proposed framework (\cf Fig.~\ref{fig:baseline}), any voice/music separation method can be combined with a \ac{hpss} method to estimate the percussive part $\hat{s}_p$
when it is not directly modeled by the \ac{bass} method (\ie KAM-REPET and RPCA). For this purpose, we simply use KAM with source-specific kernels 
\textbf{(a)+(b)} presented in Fig.~\ref{fig:kamwh}. This method is also equivalent to the median filtering approach proposed in \cite{fitzgerald2010harmonic}.
In order to enhance the harmonicity of the voice part, we can apply $F_0$ filtering on the estimated singing voice signal $\hat{s}_v$.
This method previously proposed in \cite{rpcaf0estim} for \ac{rpca}, consists in estimating at each instant the fundamental frequency $F_0$
and to apply a binary mask on a time-frequency representation to isolate the harmonic components (partials) of the predominant $F_0$ of $\hat{s}_v$, from the background music.
In our implementation, the YIN algorithm \cite{de2002yin} was used for single $F_0$ estimation before the filtering process which considers at each instant,
the spectrogram local maxima of the vicinity of each integer multiple of $F_0$, as the singing voice partials.
Hence, the residual part (not recognized as the partials) is removed from $\hat{s}_v$ and added to $\hat{s}_{h}$ (the harmonic instrumental accompaniment).
In our experiment, $F_0$ Filtering was only combined with \ac{rpca} to provide a slight improvement of the original method.

% \begin{itemize}
%  \item separation of the harmonic part from the percussive part in the mix, through median filtering (which is a particular KAM case) \cite{fitzgerald2010harmonic,fitzgerald2014harmonic},
%  \item separation of the voice from the accompaniment using KAM \cite{cho2015singing} or RPCA method \cite{candes2011robust},
%  \item Using the voice and the harmonic signal, apply a post-processing and computation of the \ac{vtmr} at each instant.
% \end{itemize}
\vspace{-0.3cm}
\subsection{Supervised Singing Voice Detection}
\subsubsection{Method description}
This technique uses a machine learning framework which remains intensively studied in the literature \cite{jamendo,regnier2009singing,schluter2016learning}.
It consists in using annotated datasets to train a classification method to automatically predict if a signal fragment of a
polyphonic music contains singing voice. Here, we propose to investigate two approaches:
%---------------------------------------
\begin{itemize}
 \item the ``classical'' supervised approach which applies singing voice detection without source separation (\ie directly on the mixture $x$),
 \item the supervised BASS approach which applies singing voice detection on the isolated signal associated to voice provided by a \ac{bass} method (\ie $\hat{s}_v$).
\end{itemize}
%---------------------------------------

%aims at reinforcing the unsupervised approach by introducing information about a singing voice through a machine learning framework.
For the classification, each signal is represented by a set of features.
In this study, we investigate separately the following descriptors: \ac{mfcc} of sources signals as proposed in \cite{jamendo},
trained KAM kernels $\mathcal{K}_i$ provided by Algorithm \ref{alg:train_kam}, \ac{ttb} \cite{timbre_toolbox} features and coefficients of
the \ac{sct} \cite{anden2011multiscale}. In order to reduce overfitting, we use the \ac{irmfsp} algorithm \cite{aes_irmfsp}
as a features selection method.

During the training step, an annotated dataset is used to model the singing voice segments and the instrumental music segments.
Hence, we obtain 3 distinct models:
\begin{itemize}
 \item when isolated voice and music signals are available (\ie MIR1K and MedleyDB), they are used to obtain respectively the models $\mu_v$ and $\mu_m$.
 \item when a singing voice is active over a music background, (\ie for all datasets) a model $\mu_{vm}$ is obtained.
\end{itemize}
During the recognition (testing) step, a trained classification method is then applied on signal fragments to detect singing voice activity.

\subsubsection{Features selection for voice detection}

In order to assess the efficiency of the proposed features for the supervised method, we computed for the Jamendo dataset \cite{jamendo},
a 3-fold cross validation (with randomly defined folds) using the \ac{svm} method with a radial basis kernel,
combined with the \ac{irmfsp} method \cite{aes_irmfsp} to obtain the top-100 best features to discriminate between vocal and musical
signal frames.
In this experiment, each music except is represented by concatenated features vectors computed on each 371~ms-long
frames (without overlap between adjacent frames). We configure each method such as \ac{kam} provides 361 values 
(using $w=h=19$), \ac{mfcc}s provide 273 values (13 MFCCS on 21 frames), \ac{ttb} provides 164 coefficients and \ac{sct} provides 866 coefficients.
The results measured in terms of F-measure are displayed in Table \ref{tab:fsvd} and shows that
\ac{sct} is the most important feature which outperforms the other ones.
Despite KAM shows its capabilities for source separation, it however provides the poorest results but close
to \ac{mfcc}s results, for singing voice detection.
The best results are obtained thanks to \ac{sct} which should be used in combination with the \ac{ttb}.
%% M_test_voice_model2
\begin{table}[!ht]
\caption{Investigation of the most efficient features for singing voice detection on the Jamendo dataset.}
\centering\resizebox{0.4\textwidth}{!}{
\begin{tabular}{c c c c | c}
 \hline
 KAM 	& MFCC & \ac{ttb} & \ac{sct} & $F_\text{meas}$\\  %& $\text{av}_\text{Rec}$ & $\text{av}_\text{Prec}$ 
 \hline
  x	&	&	&		& .75	\\ %& .74	& .74	\\
	& x	&	&		& .80	\\ %& .76	& .76	\\
	&	& x	&		& .82	\\%	&	&	\\
	&	&	& x		& \textbf{.89}	\\% 	&	&	\\    
 \hline
   x	& x	&	&		& .82	\\%	&	&	\\
   x	& 	& x	&		& .83	\\%	&	&	\\
   x	&	& 	&  x		& .88	\\ %	&	&	\\
	&  x	& x	& 		& .85	\\%	&	&	\\
	&  x	& 	&  x		& .89	\\%	&	&	\\
	&  	& x	&  x		& \textbf{.89}	\\%	&	&	\\	
 \hline
    x	& x	& x	&	& .84	\\%&	&	\\
   x	& x	& 	&  x	& .88	\\%&	&	\\
   x	&	& x	&  x	& .88	\\%&	&	\\
	&  x	& x	&  x	& \textbf{.89}	\\%&	&	\\
 \hline
   x	&  x	& x	&  x	& .89	\\%&	&	\\
\hline
\end{tabular}}
\label{tab:fsvd}
\end{table}

\section{Numerical results} \label{sec:results}
%---------------------------  M_csv_to_fig.m
\begin{figure*}[!ht]
\subfigure[RQF]{\includegraphics[width=0.32\textwidth]{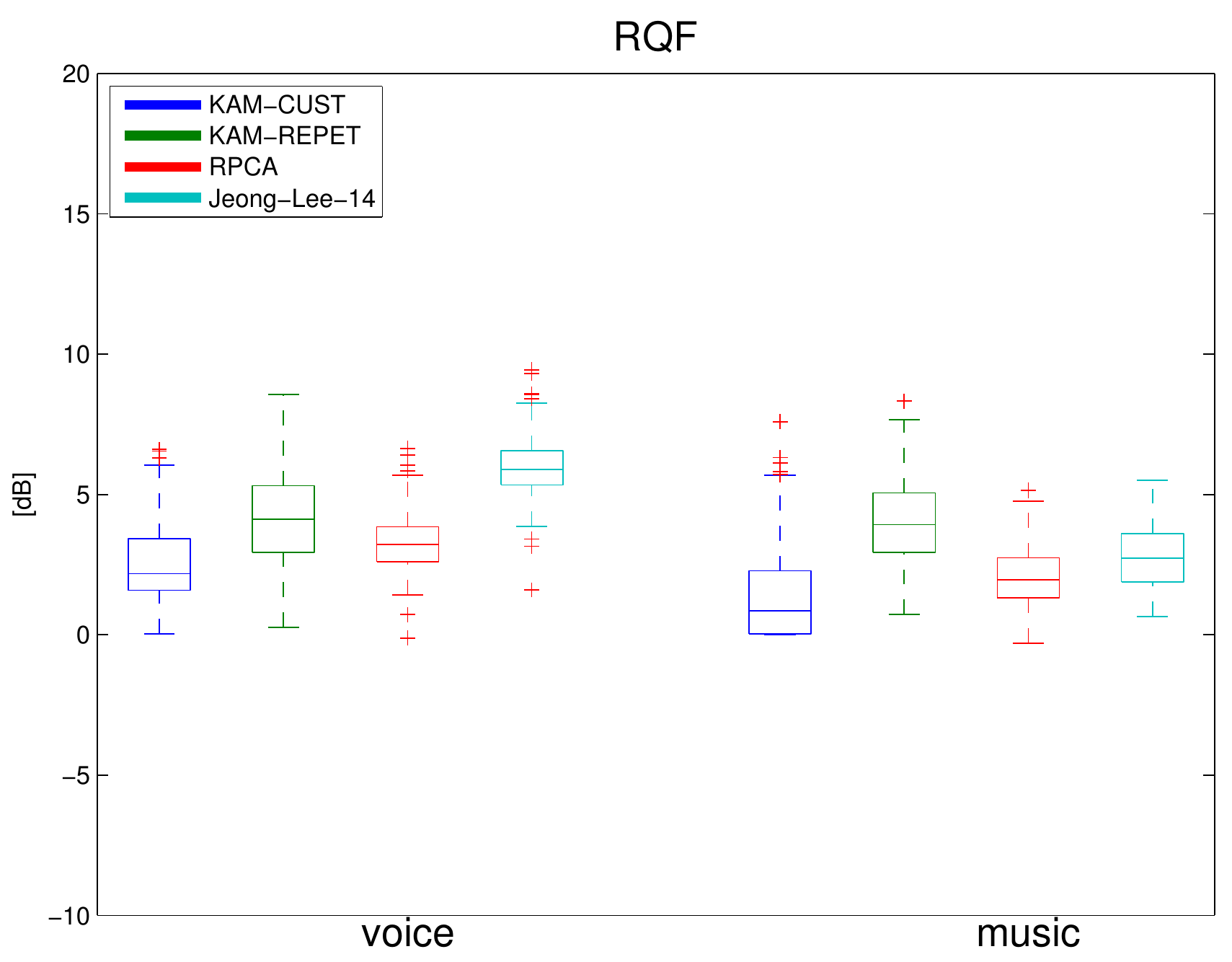}}
\subfigure[SIR]{\includegraphics[width=0.32\textwidth]{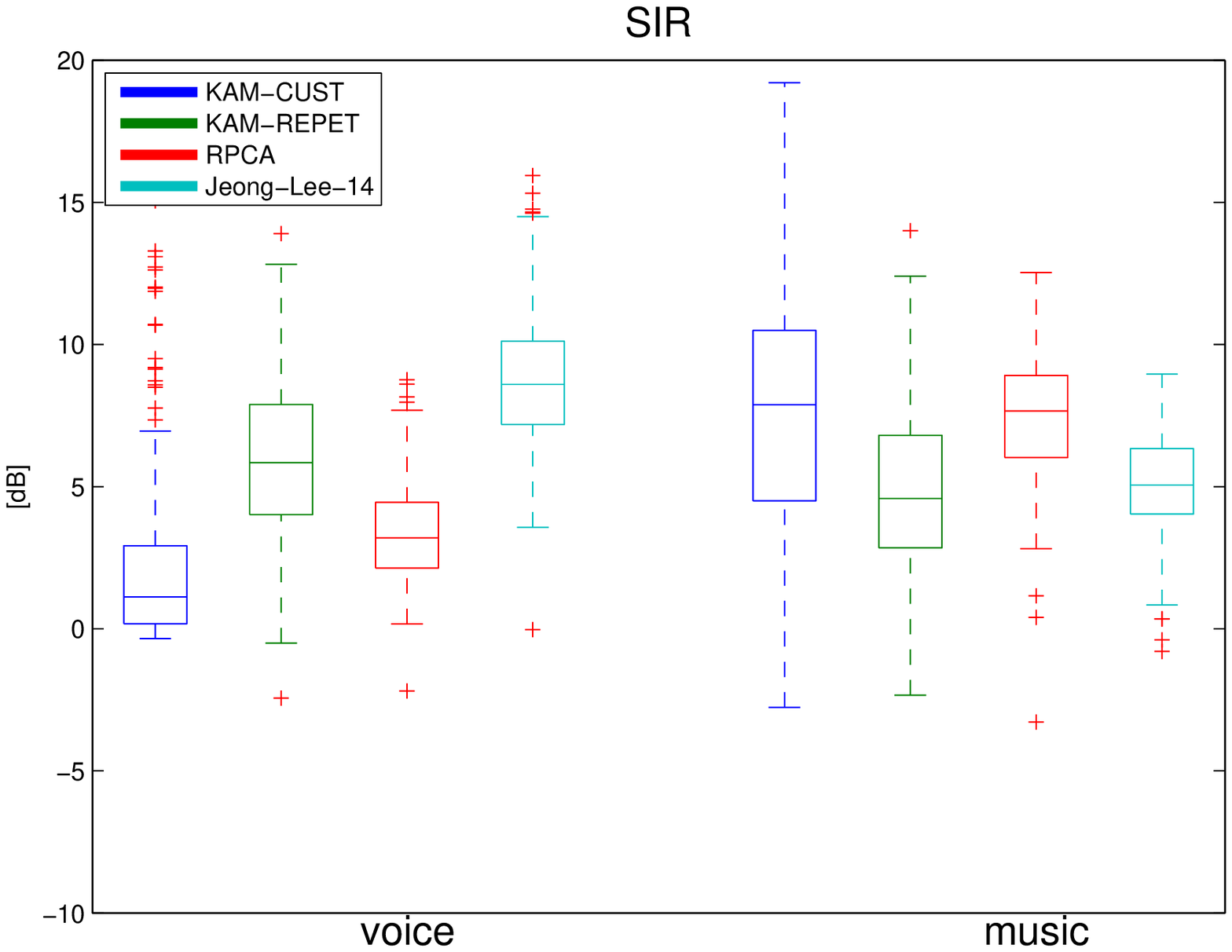}}
\subfigure[Overall-Perceptual Score (percents)]{\includegraphics[width=0.32\textwidth]{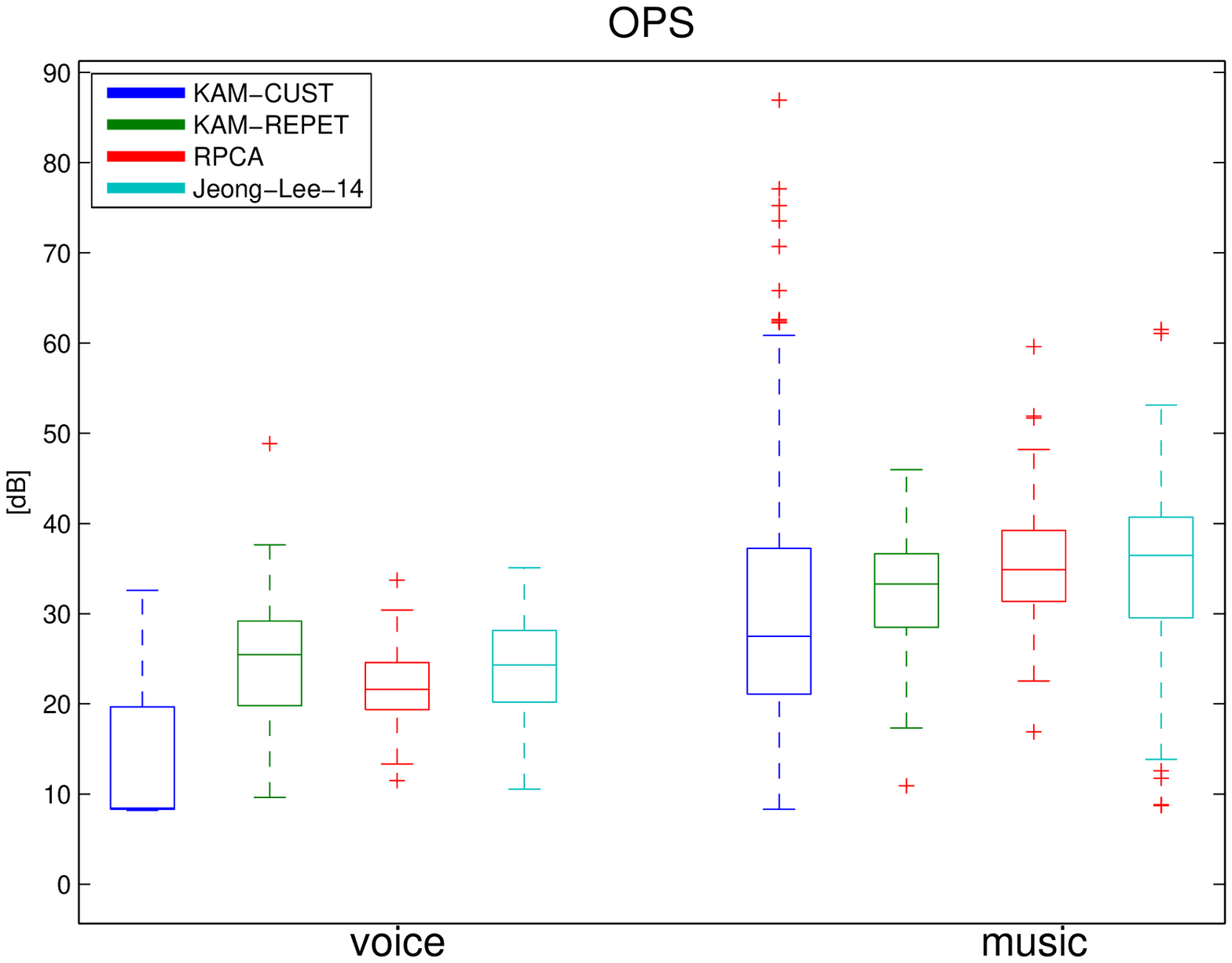}}
\caption{Objective and perceptual \ac{bass} quality results comparison on the test-fold of the MIR1K dataset.}
 \label{fig:mir1kbass}
\end{figure*}
\begin{figure*}[!ht]
\subfigure[RQF]{\includegraphics[width=0.32\textwidth]{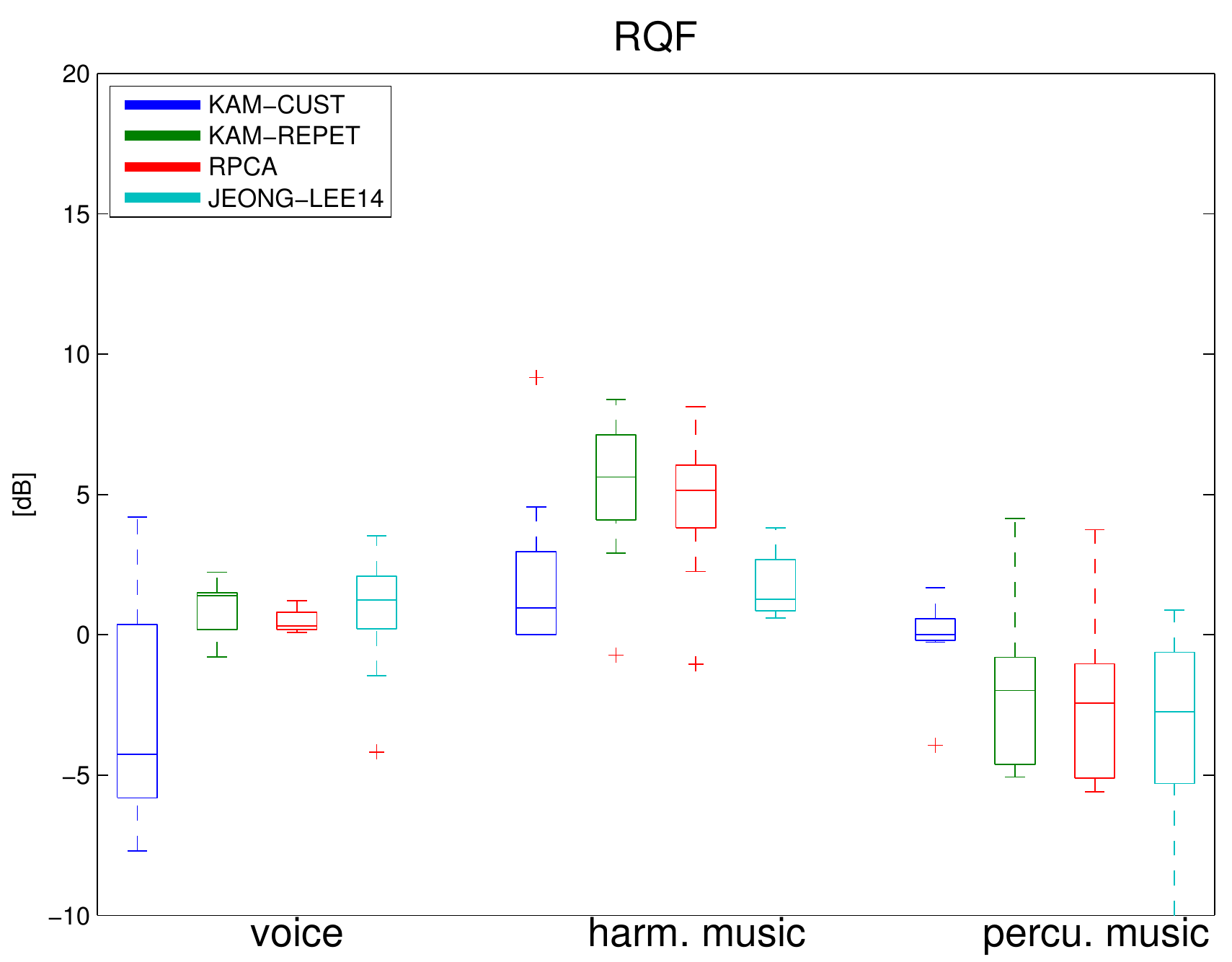}}
\subfigure[SIR]{\includegraphics[width=0.32\textwidth]{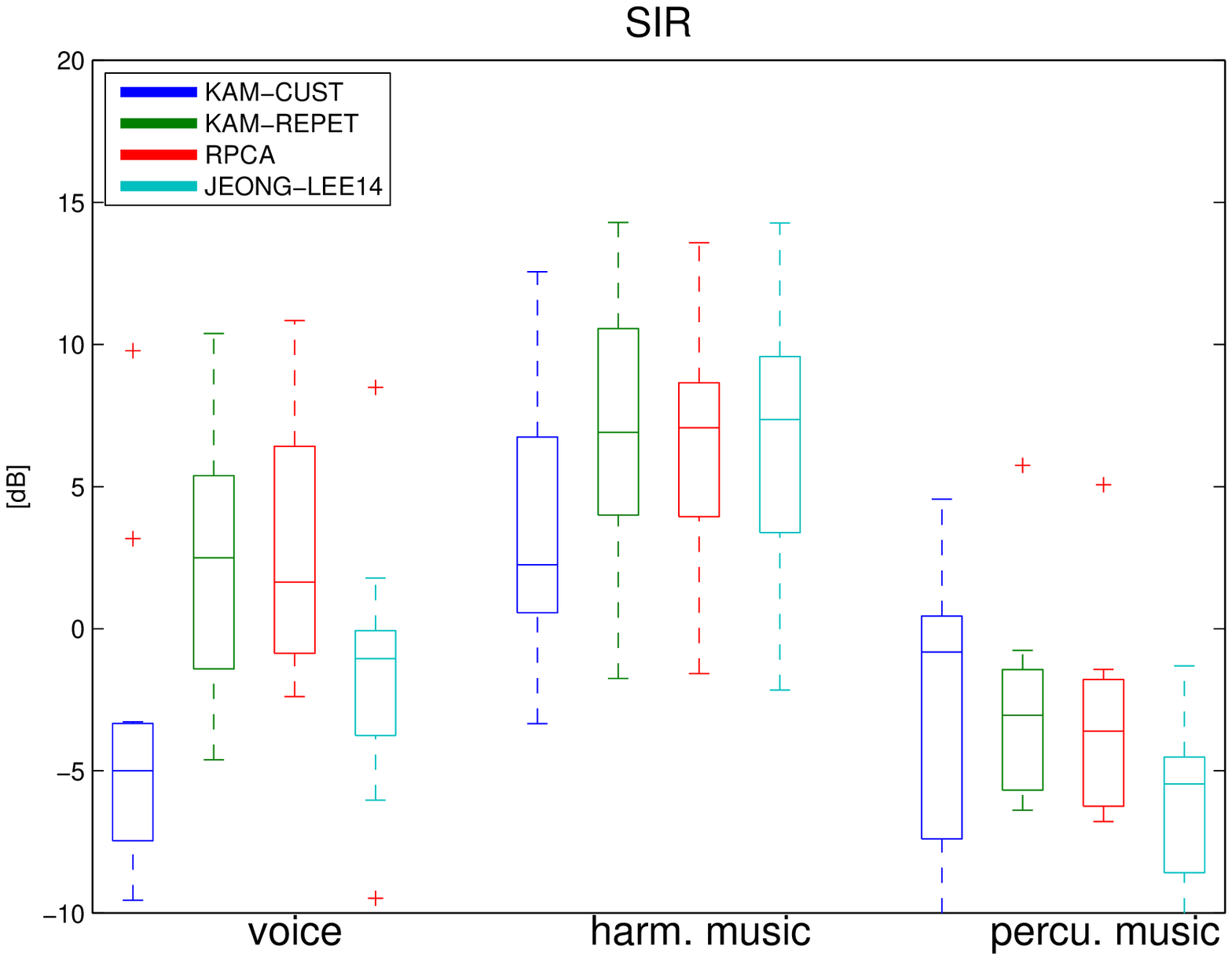}}
\subfigure[Overall-Perceptual Score (percents)]{\includegraphics[width=0.32\textwidth]{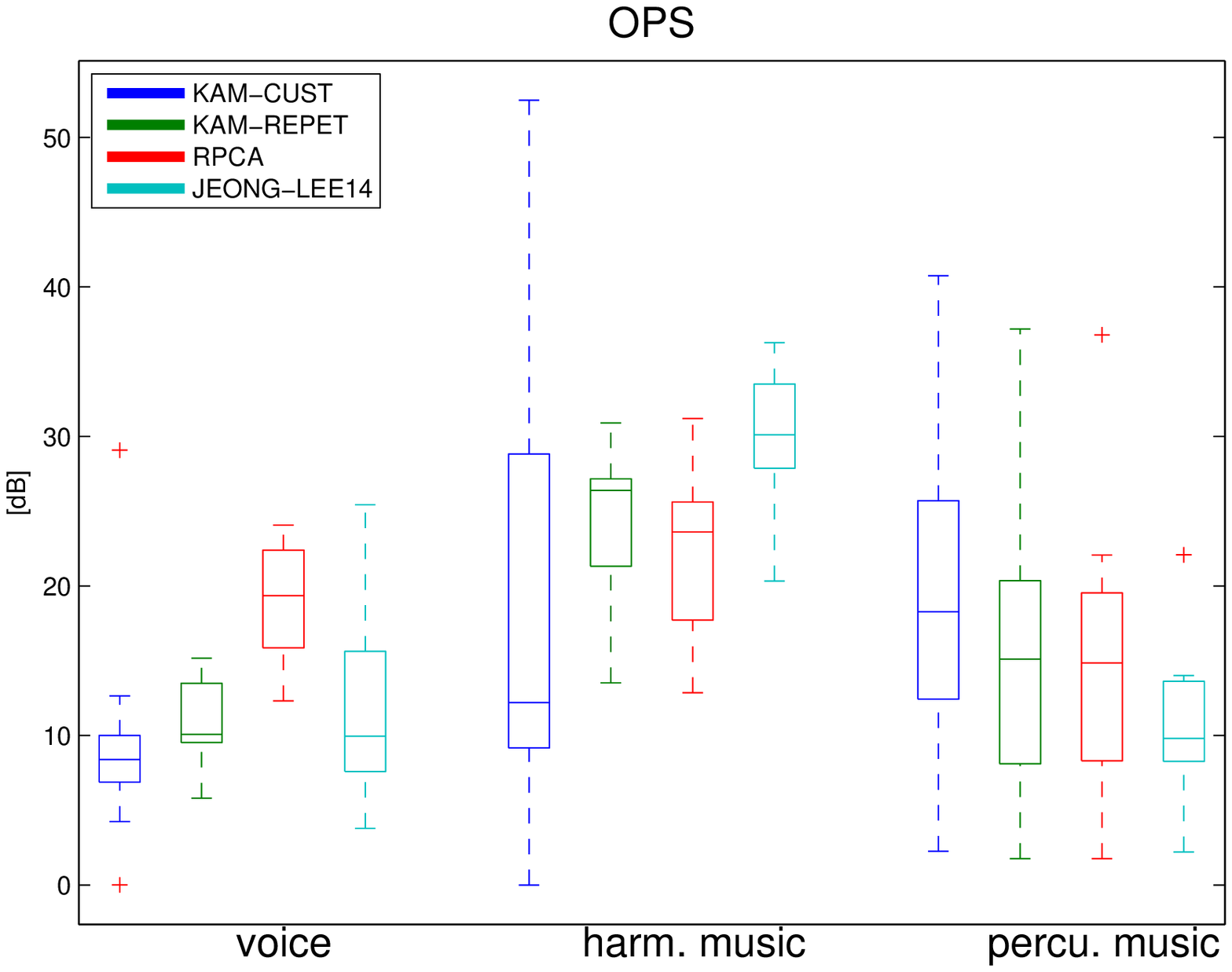}}
\caption{Objective and perceptual \ac{bass} quality results comparison on the test-fold of the MedleyDB dataset.}
 \label{fig:medleydbbass}
\end{figure*}
%--------------------------- 

\subsection{Datasets}
In our experiments, we use several common datasets allowing evaluation for source
separation (MedleyDB, MIR1K) and singing voice detection from a polyphonic mixture.
About singing voice detection, each dataset is split in several folds corresponding to training and test folds
which are both used by the evaluated supervised methods. The unsupervised methods only use the test fold.
Hence, we used 3 datasets.
\begin{itemize}
 \item Jamendo \cite{jamendo} contains creative commons music track with singing voice annotations. The whole dataset contains 93 tracks
 where 61 correspond to the training set and 16 tracks are used respectively for the test and the validation. Since the separated tracks of
 each source are not available, this dataset is only used for singing voice detection.
 \item MedleyDB \cite{MedleyDB} contains 122 music pieces of different styles, available with the separate multi-track instruments (60 with and 62 without singing voice). 
 This, allows to build a flat instantaneous single-channel mixture mix to fit the signal model proposed by Eq. \eqref{eq:mixmodel}.
 We have made a split on this dataset which preserve the ratio of voiced-unvoiced musical tracks while ensuring that each artist is only present once on each fold.
 Finally, the training dataset contains 62 tracks, the test set 36 tracks and the validation 24 tracks.
For the source separation and the singing voice detection tasks, we only focus on 50 music tracks containing singing voice.
 \item MIR1K \cite{hsu2010improvement} contains 1000 musical excerpts recorded during karaoke sessions with 19 different non-professional singers. 
 For each track the voice and the accompaniment is available.  We propose to split this dataset to obtain 828 excerpts for the training and 172 excerpts
 for the test set (containing only the singers `HeyCat' and `Amy').
\end{itemize}

\subsection{Blind Source Separation}\label{sec:results_bass}
Now, we compare the source separation performance respectively obtained on MIR1K (voice/music) and on MedleyDB (voice/music/drums) datasets
using the investigated methods: KAM-REPET, KAM-CUST, RPCA and Jeong-Lee methods. % (\cf Section \ref{sec:methods}).
For each musical track, the isolated source signals are used to construct mixtures through Eq. \eqref{eq:mixmodel}
on which the \ac{bass} methods are applied. Isolated signal are also used as references 
to compute the source separation quality measures.
Each analyzed excerpt is sampled at $\fs=22.05$~kHz and each method is configured 
to provide the best results according to Section \ref{sec:methods}: % (\cf Section \ref{sec:methods}):
\begin{itemize}
 \item KAM-REPET is a variant of the original REPET algorithm proposed
by A. Liutkus in \cite{liutkus2015} which uses a local time-varying tempo estimator to separate the
leading melody from the repetitive musical background. To obtain 3 sources (on MedleyDB), this method is combined
with the \ac{hpss} method \cite{fitzgerald2010harmonic} with $h=w=19$ (as preprocessing) to separate the percussive part.
 \item KAM-CUST is the new proposed method (\cf Section \ref{sec:kam}) based on the KAM framework using a supervised kernel training step. 
 In our experiment, we directly train the kernels on the isolated reference signals used to create the mixtures. 
 Trained kernels are configured such as $h=w=19$.
 \item RPCA corresponds to our implementation of this method with $\lambda=\frac{2}{\sqrt{\max(F,T)}}$, $\mu=10\lambda$ and $N_\text{iter}=1000$. As for the KAM-REPET method, this
 approach can be combined with the HPSS \cite{fitzgerald2010harmonic} and $F_0$-filtering to provide 2 or 3 sources when it is required.
 \item Jeong-Lee-14 corresponds to our implementation of Algorithm \ref{alg:jeong} with $\alpha\!=\!1/4$, $\phi\!=\!1/40$, $N_\text{iter}\!=\!200$, $\gamma\!=\!1/4$.
\end{itemize}
%
%\vspace{-0.2cm}
The results displayed in Fig.~\ref{fig:mir1kbass} (MIR1K) and in Fig. \ref{fig:medleydbbass} (MedleyDB)
use the boxplot representation \cite{benjamini1988opening} and measure the \ac{bass} quality in
terms of \ac{rqf}, \ac{sir} and \ac{ops} provided by BssEval$^2$ \cite{bsseval,peass}.
Jeong-Lee-14 and KAM-REPET obtain the best SIR results on MIR1K for separating the voice 
without drums separation (\cf Fig. \ref{fig:mir1kbass}). Interestingly, Jeong-Lee-14
can significantly outperforms other methods for voice separation on MIR1K, but it can also obtain the worst
results on MedleyDB.
From another side, \ac{rpca} and KAM-REPET obtain the best SIR results for separating the voice 
in combination with drums separation (\cf Fig.~\ref{fig:medleydbbass}) on MedleyDB.
Unfortunately, KAM-CUST fails to separate the voice properly.
However it can obtain the best results for accompaniment separation. This can be explained
by the variability of a singing voice spectrogram which is not sufficiently modeled by our training Algorithm.
At the contrary, better results are provided for the accompaniment which has a more stable time-frequency structure.
This can also be explained by MedleyDB for which several references signal are not
well isolated. This produces errors in the trained kernels which are used by KAM-CUST.
\footnotetext[3]{BSS Eval and PEASS: \url{http://bass-db.gforge.inria.fr/bss_eval/}}

\subsection{Singing voice detection}\label{sec:singingvoice}

Each evaluated method is configured to detect the presence of a singing voice activity 
on each signal frame of length 371.5 ms (8192 samples at $\fs=$22.05~kHz) by steps of 30 ms.
In order to compare the performance of the different proposed singing voice detection methods,
we use the recall (Rec), precision (Prec) and F-measure ($F_\text{meas}$) metrics which are commonly used 
to assess \ac{mir} systems \cite{bay2009evaluation}.
Rec (resp. Prec) is defined for each class (\ie voice ($v$) and music ($hp$)) and is averaged among classes to obtain the 
$\text{av}_\text{Rec}$ (resp. $\text{av}_\text{Prec}$).
The F-measure is thus obtained by computing the harmonic average between $\text{av}_\text{Rec}$ and $\text{av}_\text{Prec}$ such as:
% ---------------------------------------
\begin{equation}
 F_\text{meas} = 2 \frac{\text{av}_\text{Rec}\cdot \text{av}_\text{Prec}}{\text{av}_\text{Rec} + \text{av}_\text{Prec}}.
\end{equation}
% ---------------------------------------
% \begin{align}
%  \text{Rec}_{\text{inst}}	&= \frac{\hat{N}_{\text{inst}}}{N_{\text{inst}}} \\
% % \text{AvRecall} 		&= \frac{\text{Recall}_{\text{voice}} + \text{Recall}_{\text{music}}}{2}% \left( \frac{\hat{N}_{\text{voc}}}{N_{\text{voc}}} + \frac{\hat{N}_{\text{mus}}}{N_{\text{mus}}} \right) % \frac{1}{2} \left( \frac{\hat{N}_{\text{voc}}}{N_{\text{voc}}} + \frac{\hat{N}_{\text{mus}}}{N_{\text{mus}}} \right)
% \end{align}
% 
% where $\hat{N}_{\text{ins}}$ corresponds to the number of frames where the instrument $\text{inst}$ is correctly detected, and $N_{\text{ins}}$
% is the number of frames where $\text{inst}$ is present.
%\vspace{0.2cm}
\subsubsection{Unsupervised singing voice detection}

In this experiment we respectively apply the 4 investigated BASS methods described in Section \ref{sec:methods} and \ref{sec:results_bass} 
to estimate the voice source and the musical parts before applying the unsupervised approach described in Section \ref{sec:voice_unsupervised}.
Our results obtained on the MedleyDB and the MIR1K datasets are presented in Tables \ref{tab:unsupervised} (a) and (b).
The results are compared to those provided by the oracle which corresponds to the Algorithm \ref{alg:bass}
which apply a Wiener filter with $\alpha=2$ and where the isolated reference signals are assumed known.
Interestingly, the best results are reached using the KAM-REPET method without HPSS on MedleyDB
and with Jeong-Lee-14 on MIR1K with a F-measure above 0.60.

\vspace{0.2cm}
\subsubsection{BASS + supervised singing voice detection}

In this experiment, we combine a \ac{bass} method with the best \ac{svm}-based proposed supervised singing voice detection method
as investigated in Table. \ref{tab:fsvd} (\ie using TTB + SCT).
According to Tables \ref{tab:semisupervised} (a) and (b), combining \ac{bass} with supervised
singing voice detection can slightly improve the precision of detection in comparison with
the unsupervised approach (in particular KAM-REPET and KAM-CUST).
However, this approach shows a limited interest of BASS for supervised singing voice detection,
in comparison with other approaches. In fact, this approach does not allow to overcome the best score reached
through the unsupervised method, in particular the maximal recall reached for MedleyDB which remains equal to 0.59.
A solution not investigated here could be to train models specific to the results provided by a \ac{bass},
but without the insurance to obtain better results than without using \ac{bass}.

\vspace{0.2cm}
\subsubsection{Supervised singing voice detection: comparison with \ac{cnn}}

Finally, we compare all the proposed approaches (unsupervised and supervised) in terms of singing voice detection accuracy 
with an implementation of a recent state-of-the-art method \cite{schluter2016learning} based on \ac{cnn}.
The results obtained on a single dataset and after merging two datasets, are respectively displayed in Tables. \ref{tab:supervised} (a) and (b).
For the sake of clarity, we only compare the average recall results which is the most important metric.
Table \ref{tab:supervised} (b) considers two experimental cases. The Self-DB case
considers two datasets as a single dataset by merging their respective training parts (\eg MIR1K-train + JAMENDO-train)
and by merging their test parts (\eg MIR1K-test + JAMENDO-test).
%, thus the used training and the test folds correspond to the fusion of the corresponding folds of the merged datasets.
The cross-DB case uses two merged datasets for the training step (\eg MIR1K-train + JAMENDO-train)
and uses the third dataset for testing the singing voice detection (\ie MedleyDB-test).
Results show that the \ac{cnn}-based method outperforms the proposed unsupervised and the supervised methods when
it is applied on single datasets (\cf (a) and seld-DB (b)).
However, the unsupervised approach can beat \ac{cnn} in cross-DB (b) case.
This is visible for the MIR1K where the best unsupervised methods (RPCA and Jeong-Lee-14) obtain a recall equal to 0.68
when the \ac{cnn}-based method is trained on Jamendo+MedleyDB only 0.65. 
This result shows that an unsupervised approach can also be of interest to avoid overfitting or when no training
dataset is available. Moreover, our proposed supervised methods can obtain comparable results to \ac{cnn} in the cross-DB case
except for singing voice detection applied on MIR1K.

%% JAMEDO
%%	   	rec_v		rec_m		av_rec		prec_v		prec_m		av_prec		f-meas
%Jeon14:   	0.1611875	0.8980625	0.5295625	0.6175		0.508125	0.5628125	0.525875
%KAM1: 		0.2255		0.9478125	0.5868125	0.8174375	0.480875	0.649125	0.598375
%RPCA:		0.0455625	0.9855		0.5155		0.7363076923	0.51625		0.6176153846	0.5435384615

%% MEDLEYDB
%Jeon14:	0.1370588235	0.8876315789	0.5088823529	0.6787894737	0.6231578947	0.6507894737	0.5903529412
%KAM1		0.3059411765	0.8892631579	0.5945294118	0.7801578947	0.5850526316	0.6826315789	0.6582352941
%KAM2		0.3635294118	0.7288421053	0.5331176471	0.5965833333	0.5040526316	0.4785833333	0.538
%RPCA:		0.0607058824	0.9851052632	0.5231176471	0.8747058824	0.6173157895	0.7607647059	0.6328125

\begin{table}[!ht]
\caption{Unsupervised voice detection results using \ac{bass} (bold values denotes best results except for Oracle).}
\centering
\subfigure[with and without drums separation on the MedleyDB dataset]
{
\resizebox{0.45\textwidth}{!}{
\begin{tabular}{|c||c|c|c|}
 \hline
						&  av. Rec.	&   av. Prec. 	& F-meas \\
 \hline
 \hline
 Oracle						& 0.71		&  	0.66	& 0.68  \\
 \hline
 KAM-REPET					& \textbf{0.59}	&	0.68	& \textbf{0.63}	 \\
 \hline
 KAM-REPET + HPSS				& 0.54		&	0.69	& 0.60	 \\
 \hline
 KAM-CUST					& 0.50		&  	0.62	& 0.55	 \\
 \hline
 \ac{rpca}					& 0.52		&\textbf{0.76}	& 0.61	\\
 \hline
 \ac{rpca} + HPSS				& 0.53		& 	0.75	& 0.62	 \\
 \hline
 Jeong-Lee-14					& 0.50		&	0.65	& 0.56	 \\
 \hline
 \end{tabular}}}\\
 \subfigure[without drums separation applied on the MIR1K dataset]{
\resizebox{0.45\textwidth}{!}{
\begin{tabular}{|c||c|c|c|}
\hline
						&  av. Rec.	& av. Prec. 	& F-meas\\
 \hline
 \hline
 Oracle						& 0.82		&	0.72	& 0.76  \\
 \hline
 KAM-REPET					& 0.65		&  	0.75	& 0.69  \\
 \hline
 KAM-CUST					& 0.57		&  	0.55	& 0.55  \\
 \hline
 \ac{rpca}					& \textbf{0.68}	& 	0.61	& 0.64  \\
 \hline
 Jeong-Lee-14					& \textbf{0.68}	& \textbf{0.78}	& \textbf{0.72}	\\
 \hline
 \end{tabular}}}
 \label{tab:unsupervised}
\end{table}

%%%%%%%%%%%%%%%%%%%%%%%%%%%%%%%%%%%%%%%%%%%%%%%% BASS + VD
\begin{table}[!ht]
\caption{\ac{bass} combined with supervised singing voice detection results (bold values denotes best results except for Oracle).}
\centering
\subfigure[with drums separation applied on the MedleyDB dataset]{
\resizebox{0.45\textwidth}{!}{
\begin{tabular}{|c||c|c|c|}
\hline
						& av. Rec.	& av. Prec. 	& F-meas\\
\hline
\hline
 Oracle						& 0.71		& 0.68		& 0.69	\\
 \hline
 KAM-REPET + HPSS				& 0.52		& \textbf{0.76}	& \textbf{0.61}	\\
 \hline
 KAM-CUST					& \textbf{0.59}	& 0.64		& \textbf{0.61}	\\
 \hline
 \ac{rpca} + HPSS				& 0.55		& 0.69		& \textbf{0.61}	\\
 \hline
 Jeong-Lee-14					& 0.49		& 0.64		& 0.55	\\
 \hline
 \end{tabular}}}\\
 \subfigure[without drums separation applied on the MIR1K dataset]{
 \centering
\resizebox{0.45\textwidth}{!}{
\begin{tabular}{|c||c|c|c|}
\hline
						&  av. Rec.	&  av. Prec. 	& F-meas\\
\hline
\hline
 Oracle						& 0.67		&	0.61	& 0.63	\\
 \hline
 KAM-REPET					& \textbf{0.60}	& 	0.70	& \textbf{0.64}	\\
 \hline
 KAM-CUST					& 0.52		& 	0.62	& 0.56	\\
 \hline
 \ac{rpca}					& 0.55		& \textbf{0.74} & 0.63	\\
 \hline
 Jeong-Lee-14					& 0.51		& 	0.72	& 0.59	\\
 \hline
 \end{tabular}}}
 \label{tab:semisupervised}
\end{table}
%%%%%%%%%%%%%%%%%%%%%%%%%%%%%%%%%%%%%%%%%%%%%%%%%%%%%%%%%%%%%%%%%%%%%%%%%%%%%

\begin{table}[!ht]
\caption{Comparison of the proposed methods with \cite{schluter2016learning} measured in terms of average recall for singing voice detection.}
\subfigure[evaluation on each dataset]{
\centering
\resizebox{0.45\textwidth}{!}{
\begin{tabular}{c || c | c  | c}
Dataset 	& Best unsupervised 			& SVM (MFCC+SCT)	& CNN \\
\hline
Jamendo 	&	0.58				& 	0.81		& 0.86 \\  % (0.89 with bagging)
MIR1K 		& 	0.68				&	0.77		& 0.9 \\
MedleyDB 	&	0.59				& 	0.79		& 0.86 \\
\end{tabular}}}\\
\subfigure[evaluation on merged datasets]{
\resizebox{0.45\textwidth}{!}{
\begin{tabular}{c || c  c | c  c}
Training datasets 		& \multicolumn{2}{c}{SVM (MFCC+SCT)} & \multicolumn{2}{|c}{CNN} \\
				& self-DB 		& cross-DB  				& self-DB  		& cross-DB \\
\hline
Jamendo + MIR1K  		& 	0.81		& 	0.73				& 0.89 			& 0.75 \\
Jamendo + MedleyDB 		&	0.80		&	0.59				& 0.86 			& 0.65\\
MedleyDB + MIR1K 		&	0.80		&	0.76				& 0.84 			& 0.77 \\
\end{tabular}}
}
\label{tab:supervised}
\end{table}

\section{Conclusion} \label{sec:conclusion}

We have presented recent developments for blind single-channel audio source separation methods,
which use morphological filtering of the mixture spectrogram.
These methods were compared together for source separation and using our new framework for singing voice detection 
which uses \ac{bass} as a preprocessing step. We have also proposed a new contribution to extend the KAM framework to automatically design 
kernels which fits any given audio source.
Our results show that our proposed KAM-CUST method is promising and
can obtain better results than KAM-REPET for blind source separation.
However, our training algorithm is sensitive and should
be further investigated to provide discriminative source-specific kernels.
Moreover, we have shown that the unsupervised approach remains of interest
for singing voice detection in comparison with more efficient method such as
\cite{schluter2016learning} based on \ac{cnn}.
In fact, the weakness of supervised approaches can become visible when large
databases are processed or when a few annotated examples are available.
Hence, this study paves the way of a future investigation of the KAM framework 
in order to efficiently design source-specific kernels which can be used both for source separation or for
singing voice detection. Future works will consider new practical applications of the proposed methods
while improving the robustness of the new proposed KAM training algorithm.

\section*{Acknowledgement}

This research has received funding from the European Union's Horizon 2020 research 
and innovation program under grant agreement n$^o$ 688122.
Thanks go to Alice Cohenhadria for her implementation of method \cite{schluter2016learning}
used in our comparative study.

%\ifCLASSOPTIONcaptionsoff
%  \newpage
%\fi



%\bibliographystyle{IEEEtran}
\bibliographystyle{IEEEtran}

%\bibliography{IEEEabrv,biblio}

% \end{IEEEbiography}
% % insert where needed to balance the two columns on the last page with
% %\newpage


%\vfill


\end{document}